\documentclass[aps,pra]{revtex4-2}
\usepackage{amsfonts}
\usepackage{amsmath}
\usepackage{amsthm}
\usepackage{amssymb}
\usepackage{graphicx}
\usepackage{subfigure}

\begin{document}

\title{Past and present trends in the development of the pattern-formation
theory: domain walls and quasicrystals }
\author{Boris A. Malomed$^{1.2}$}
\affiliation{$^{1}$Department of Physical Electronics, School of Electrical Engineering,
Faculty of Engineering, and Center for Light-Matter Interaction, Tel Aviv
University, P.O. Box 39040 Tel Aviv, Israel\\
$^{2}$Instituto de Alta Investigaci\'{o}n, Universidad de Tarapac\'{a}, Casilla 7D,
Arica, Chile}

\begin{abstract}
A condensed review is presented for two basic topics in the theory of
pattern formation in nonlinear dissipative media: (i) domain walls (DWs,
alias grain boundaries), which appear as transient layers between different
states occupying semi-infinite regions, and (ii) two- and three-dimensional
(2D and 3D) quasiperiodic (QP) patterns, which are built as superposition
of plane-wave modes with incommensurate spatial periodicities. These topics
are selected for the present article, dedicated to the 70th birthday of
Professor M. I. Tribelsky, due to the impact made on them by papers
published by Prof. Tribelsky and his coauthors. Although some findings
revealed in those works may now seem as \textquotedblleft old" ones, they
keep their significance as fundamentally important results in the the theory
of nonlinear DW and QP patterns. Adding to the findings revealed in the
original works by M. I. Tribelsky \textit{et al}., the present article also
reports several new analytical results, obtained as exact solutions to
systems of coupled real Ginzburg-Landau (GL) equations. These are: a new
solution for symmetric DWs in the bimodal system including linear mixing
between its components; a solution for a strongly asymmetric DWs in the case
when the diffusion (second-derivative) term is present only in one GL
equation; a solution for a system of three real GL equations, for the
symmetric DW with a trapped bright soliton in the third component; and an
exact solution for DWs between counter-propagating waves governed by the GL
equations with group-velocity terms. The significance to the
\textquotedblleft old" and new results collected in this review is enhanced
by the fact that the systems of coupled equations for two- and
multicomponent order parameters, addressed in the article, apply equally
well to modeling thermal convection, multimode light propagation in
nonlinear optics, and binary Bose-Einstein condensates.
\end{abstract}

\maketitle

\noindent \textbf{Dedicated to the celebration of the 70th birthday of
Professor Mikhail Isaakovich Tribelsky}

\section{Introduction}

\subsection{The objective of this article}

Apart from his fundamental contributions to optics, especially to the theory
of the nonresonant light-matter interaction \cite{Anisimov,MIT1} and light
scattering by small particles \cite{MIT1}-\cite{MIT6}, an essential topic in
the works of Mikhail Tribelsky \cite{name} has been the theory of pattern
formation in nonlinear dissipative media. In particular, two important
subjects considered in his publications are domain walls (DWs, alias \textit{%
grain boundaries}), i.e., stationary stripes separating two domains which
are filled by different stable patterns, and quasiperiodic (QP) patterns,
alias dissipative two-dimensional (2D) quasicrystals. It is relevant to
mention that the fundamental papers of Prof. Tribelsky on the former and
latter topics, \textit{viz}., Refs. \cite{Trib-DW} and \cite{Trib-quasi},
are, respectively, his second and sixth best-cited publications, according
to the data provided by \textit{Web of Science}. The objective of this
article is to produce a condensed review of basic results reported in those
old but still significant works, and outline directions of subsequent work
which was initiated by results reported in them. The article also includes
some new exact analytical results for DWs which offer a natural extension of
the analysis initiated in Ref. \cite{Trib-DW} (in a detailed form, the new
results will be reported elsewhere \cite{new}). The presentation given in
the article has a personal flavor, due to the fact that the present author
was a Mikhail's collaborator in projects which had produced the
above-mentioned original publications.

In addition Ref. \cite{Trib-DW}, it is relevant to mentioned a still earlier
paper \cite{early}, where we addressed a well-known model equation, which is
usually called the real Ginzburg-Landau (GL) equation (the name originates
from the phenomenological theory of superconductivity elaborated by Ginzburg
and Landau 70 years ago \cite{GiLa}). The usual scaled form of the real GL
equation is%
\begin{equation}
\frac{\partial u}{\partial t}=u+\frac{\partial ^{2}u}{\partial x^{2}}%
-|u|^{2}u.  \label{realGL}
\end{equation}%
Actually, the \textit{order parameter} $u(x,t)$ governed by Eq. (\ref{realGL}%
) is a complex function, while the equation is called \textquotedblleft
real" because its coefficients are real (therefore, by means of scaling, all
coefficients in Eq. (\ref{realGL}) are set to be $\pm 1$). The first,
second, and third terms on the right-hand side of Eq. (\ref{realGL})
represent, respectively, the linear gain, diffusion/viscosity (dispersive
linear loss), and nonlinear loss. The real GL equation is a universal model
for many nonlinear dissipative media, such as the Rayleigh-B\'{e}nard
convective instability in a shallow layer of a liquid heated from below \cite%
{Cross1982} and instability of a plane laser evaporation front \cite%
{Anisimov}.

Note that Eq. (\ref{realGL}) may be represented in the \textit{gradient form}%
,%
\begin{equation}
\frac{\partial u}{\partial t}=-\frac{\delta L}{\delta u^{\ast }},
\label{grad}
\end{equation}%
where $\delta /\delta u^{\ast }$ stands for the functional (Frech\'{e})
derivative, and
\begin{equation}
L=\int_{-\infty }^{+\infty }\left( -|u|^{2}+\left\vert \frac{\partial u}{%
\partial x}\right\vert ^{2}+\frac{1}{2}|u|^{4}\right) dx  \label{Lyapunov}
\end{equation}%
is a real \textit{Lyapunov functional}. A consequence of the gradient
representation is that $L$ may only decrease or stay constant in the course
of the evolution, $dL/dt\leq 0$. This fact simplifies the dynamics of the
real GL equation.

This equation gives rise to a family of simple stationary plane-wave (PW)
solutions,%
\begin{equation}
u(x)=\sqrt{1-k^{2}}\exp \left( ikx\right) ,  \label{CW}
\end{equation}%
where real wavenumber $k$ takes values in the existence band,%
\begin{equation}
-1<k<+1.  \label{-1+1}
\end{equation}%
A nontrivial issue is stability of the PW solutions against small
perturbations. It can be naturally addressed by rewriting Eq. (\ref{realGL})
in the \textit{Madelung form} (sometimes called the hydrodynamic
representation), substituting
\begin{equation}
u\left( x,t\right) =A\left( x,t\right) \exp \left( i\phi \left( x,t\right)
\right) ,  \label{Madelung}
\end{equation}%
where $A$ and $\phi $ are real amplitude and phase. The substitution splits
Eq. (\ref{realGL}) into a pair of real equations:%
\begin{gather}
\frac{\partial A}{\partial t}=A+\frac{\partial ^{2}A}{\partial x^{2}}%
-A\left( \frac{\partial \phi }{\partial x}\right) ^{2}-A^{3},
\label{Madelung-A} \\
A\frac{\partial \phi }{\partial t}=A\frac{\partial ^{2}\phi }{\partial x}+2%
\frac{\partial A}{\partial x}\frac{\partial \phi }{\partial x}.
\label{Madelung-phi}
\end{gather}

In terms of this system, the PW solution (\ref{CW}) is written as
\begin{equation}
A=\sqrt{1-k^{2}},\phi (x)=kx.  \label{Aphi}
\end{equation}%
In paper \cite{early} (incidentally, it is the ninth best-cited publication
of M. I. Tribelsky), the stability of solution (\ref{Aphi}) was explored by
means of linearization of Eqs. (\ref{Madelung-A}) and (\ref{Madelung-phi})
against small perturbations of the amplitude and phase. We had thus found
that the stability region in the existence band (\ref{-1+1}) is
\begin{equation}
-1/\sqrt{3}\leq k\leq +1/\sqrt{3}.  \label{Eck}
\end{equation}%
In this region, the squared amplitude of the PW solution, $A^{2}(k)$,
exceeds $2/3$ of its maximum value, $A_{\max }^{2}\equiv 1$, which
corresponds to $k=0$:%
\begin{equation}
A^{2}\equiv 1-k^{2}\geq 2/3.  \label{2/3}
\end{equation}

At that time, we were not aware of the fact that this result, in the form of
Eq. (\ref{Eck}) was established much earlier \cite{Eckhaus} by Wiktor
Eckhaus \cite{Wiktor} It is now commonly known as the \textit{Eckhaus
stability criterion} (ESC). Later, we had learned that some other people
entering this research area had also independently rediscovered the ESC.
This fact had suggested our coauthor in Refs. \cite{Trib-DW} and \cite%
{Trib-quasi}, Prof. Alexander Nepomnyashchy, to formulate a \textit{%
Nepomnyashchy criterion}: a necessary condition for successful work on the
pattern-formation theory is the ability of the researcher to re-derive the
ESC from the scratch.

\subsection{Complex Ginzburg-Landau equations: the formulation, plane waves,
and dissipative solitons}

Before proceeding to the discussion of particular topics included in this
article, it is relevant to briefly recapitulate the main principles
concerning complex GL equations, as a class of fundamental models underlying
the theory of pattern formation under the combined action of linear gain and
loss (including the diffusion/viscosity), linear wave dispersion, nonlinear
loss, and nonlinear dispersion. In the case of the cubic nonlinearity, the
generic form of this equation is \cite{AransonKramer,Encycl}
\begin{equation}
\frac{\partial u}{\partial t}=gu+\left( a+ib\right) \frac{\partial ^{2}u}{%
\partial x^{2}}-\left( d+ic\right) |u|^{2}u,  \label{CCGL}
\end{equation}%
cf. its counterpart (\ref{realGL}) with real coefficients. Here, constants $%
g>0$, $a\geq 0$, and $d>0$ represents, severally, the linear gain, diffusion
coefficient, and nonlinear loss. Further, coefficients $b$ and $c$, which
may have any sign, control the linear and nonlinear dispersion,
respectively. Coefficient $g$ in Eq. (CCGL) may include an imaginary part
too, but such a frequency term can be trivially removed by a transformation,
$u(x,t)\equiv u(x,t)\exp \left[ i\mathrm{Im}(g)t\right] $.

By means of obvious rescaling of $t$, $x$, and $u$, one can fix three
coefficients in Eq. (\ref{CCGL}):%
\begin{equation}
g=a=d=1,  \label{11}
\end{equation}%
unless the equation does not include the diffusion term, in which case $a=0$
is set. Equation (\ref{CCGL}) is written in the 1D form, while its
multidimensional version is obtained replacing $\partial ^{2}u/\partial
x^{2} $ by the Laplacian, $\nabla ^{2}u$.

Unlike Eq. (\ref{realGL}), the complex GL equation (\ref{CCGL}) does not
admit a gradient representation (see Eq. (\ref{grad})). In the case of
relatively small real parts of the coefficients, i.e., $a\ll |b|$, $d\ll |c|$%
, Eq. (\ref{CCGL}) may be treated as a perturbed version of the nonlinear
Schr\"{o}dinger (NLS) equation. Methods of the perturbation theory for NLS
equations had been developed in detail long ago \cite{KM}.

The ubiquity of the complex GL equations is stressed by the title of the
major review by Aranson and Kramer \cite{AransonKramer}, \textquotedblleft
The world of the complex Ginzburg-Landau equation" -- indeed, the great
number of particular forms of such equations, their various realizations and
applications, and the great number of solutions, obtained by means of
numerical and approximate analytical methods, form a \textquotedblleft
world" in itself. As concerns applications, complex GL equations emerge not
only in areas, such as optics of laser cavities \cite%
{Arrechi,Rosanov1,Rosanov2,Lega}, where they can be directly derived as
basic physical models, with $u(x,t)$ being a slowly-varying amplitude of the
optical field, but also in many other areas of physics (hydrodynamics,
electron-hole plasmas in semiconductors and gas-discharge plasmas, chemical
waves, etc.). In many cases, underlying systems of basic equations are
complicated, but complex GL equations may be derived from them as asymptotic
equations for long-scale small-amplitude (but, nevertheless, essentially
nonlinear) excitations \cite{CrossHohenberg,Kramer,Hoyle}. In some cases,
equations of the complex-GL type may also be quite useful as
phenomenological models \cite{AransonKramer,Encycl}.

While DW states are supported by a finite-amplitude PW\ background, it is
relevant to mention that complex GL equations may give rise to localized
states (dissipative solitons \cite{CQ-first}-\cite{Tang}). In particular,
Eq. (\ref{CCGL}) admits an exact solution,%
\begin{equation}
u=A\,\left[ \cosh \left( \kappa x\right) \right] ^{-\left( 1+i\mu \right)
}\exp \left( -i\omega t\right) ,  \label{Stenflo}
\end{equation}%
with a single set of parameter values, $A$, $\kappa $, $\mu $, and $\omega $%
, given by cumbersome expressions \cite{Stewartson,Lennart}. If the complex
GL equation reduces to a perturbed NLS equation, the dissipative soliton (%
\ref{Stenflo}) can be obtained from the NLS soliton by means of the
perturbation theory, under condition $bc<0$ (otherwise, the underlying NLS
equation does not have bright-soliton solutions). However, solution (\ref%
{Stenflo}) is always unstable, as the linear gain in Eq. (\ref{CCGL}),
represented by $g>0$, makes the zero background around the soliton unstable.

Dissipative solitons of this type may be effectively stabilized, in a
nonstationary form, in a model including time-periodic alternation of linear
gain and loss, which implies replacing the constant coefficient $g$ in Eq. (%
\ref{CCGL}) by function $g(t)$ periodically changing between positive and
negative values; in particular, it may be taken as a periodic array of
amplification pulses on top of a constant lossy background,%
\begin{equation}
g(t)=G\sum_{n=-\infty }^{+\infty }\delta \left( t-\tau n\right) -g_{0},
\label{g}
\end{equation}%
with $G>0$ and $g_{0}>0$, $\tau $ being amplification period \cite{JOSA-B}.
Another option for the stabilization is the use of the \textit{dispersion
management}, i.e., replacing constant dispersion coefficient $b$ in Eq. (%
\ref{CCGL}) by function $b(t)$ which periodically jumps between positive
and negative values, cf. Eq. (\ref{g}) \cite{Berntson}.

The fact that the dissipative soliton (\ref{Stenflo}) may be considered as
an extension of bright NLS solitons suggests that Eq. (\ref{CCGL}) may also
support a solution resembling the dark soliton of the NLS equation with the
self-defocusing nonlinearity. Indeed, such solutions were found buy Nozaki
and Bekki in the form of ``holes" \cite{holes}. Although the holes, as well
as DWs, are supported by a stable PW background, they are completely
different states, as DWs separate \emph{different} PWs (see below), while
the hole is built into a single PW.

A more sophisticated version of the complex GL equation admits the existence
of \emph{stable} stationary dissipative solitons, if the zero solution is
stable, i.e., the linear term must be lossy, corresponding to $g<0$ in Eq. (%
\ref{CCGL}). In this case, it is necessary to include the cubic gain and
quintic loss (the latter term prevents the blowup). Thus, one arrives at the
complex GL equation with the cubic-quintic nonlinearity, which was first
introduced by Petviashvili and Sergeev \cite{Petviashvili} (actually, as a
2D equation) in the form of
\begin{equation}
\frac{\partial u}{\partial t}=gu+\left( a+ib\right) \frac{\partial ^{2}u}{%
\partial x^{2}}-\left( d+ic\right) |u|^{2}u-\left( f+ih\right) |u|^{4}u,
\label{CQ}
\end{equation}%
with $g<0$, $a\geq 0$, $d<0$, and $f>0$, cf. Eq. (\ref{CCGL}).

It follows from Eq. (\ref{CQ}) that the interplay of the gain and loss terms
in Eq. (\ref{CQ}) allows generation of nonzero states under the condition
that the cubic-gain strength exceeds a minimum value necessary to compensate
the effect of the loss:
\begin{equation}
|d|>\left( |d|\right) _{\min }=2\sqrt{|g|f}.  \label{dgf}
\end{equation}%
Further, using the rescaling freedom, one can normalize Eq. (\ref{CQ}) by
setting%
\begin{equation}
-g=a=-d\equiv 1,  \label{1}
\end{equation}%
cf. Eq. (\ref{11}) in the case of Eq. (\ref{CCGL}). Then, condition (\ref%
{dgf}) amounts to $f<1/4$.

Stable dissipative solitons as solutions of Eq. (\ref{CQ}), in the case when
they may be considered as a perturbation of the NLS solitons, were first
predicted in Ref. \cite{CQ-first}, and later rediscovered in Ref. \cite%
{Fauve}. Then, stable dissipative-soliton solutions of Eq. (\ref{CQ}) were
found in the opposite limit, when the dispersive terms in this equation may
be treated as small perturbations. In this case, the dissipative solitons
are broad (nearly flat) states, bounded by sharp edges in the form of a kink
and antikink \cite{Hohenberg,Hakim,Alik}.

The complex GL equation (\ref{CQ}), subject to normalization (\ref{1}),
generates a family of PW solutions, where the wavenumber takes values in the
same interval (\ref{-1+1}) as above:
\begin{equation}
\psi =\sqrt{1-k^{2}}\exp \left( ikx-i\omega t\right) ,\,\omega =c+(b-c)k^{2},
\label{PW}
\end{equation}%
cf. stationary solutions (\ref{CW}) of the real GL equation. The stability
of these flat states against long-wave perturbations can be investigated
analytically, leading to a generalization of the ESC (cf. Eq. (\ref{Eck})):%
\begin{equation}
k^{2}\leq \left( 1+bc\right) /\left( 3+2c^{2}+bc\right)  \label{long-wave}
\end{equation}%
\cite{Tsuzuki}. The full stability of solutions (\ref{PW}) was investigated
in a numerical form \cite{AransonKramer,Encycl}. Note that condition (\ref%
{long-wave}) cannot hold unless the dispersion coefficients in Eq. (\ref{CQ}%
), normalized as per Eq. (\ref{1}), satisfy the \textit{Benjamin-Feir-Newell}
(BFN)\ condition,%
\begin{equation}
1+bc>0.  \label{BFN}
\end{equation}%
If this condition does not hold, unstable PWs develop \textit{phase
turbulence}, with $|\psi |$ staying roughly constant, while the phase of the
complex order parameter, $\phi (x,t)\equiv \arg \left\{ u(x,t)\right\} $,
demonstrates spatiotemporal chaos. Just below the BFN instability threshold,
i.e., at $0<-\left( 1+bc\right) \ll 1$ (see Eq. (\ref{BFN})), the chaotic
evolution of the phase gradient $p\equiv \phi _{x}$ obeys the \textit{%
Kuramoto-Sivashinsky equation} \cite{Siva,Kura}, whose scaled form is
\begin{equation}
p_{t}+p_{xx}+p_{xxxx}+pp_{x}=0.  \label{KS}
\end{equation}%
Deeper into the region of $1+bc<0$, the instability creates \textit{defects}
of the wave field, at which $|u\left( x,t\right) |=0$, and eventually leads
to the onset of \textit{defect turbulence} \cite{AransonKramer,Encycl}.
Further evolution may lead to emergence of regularly arranged train-shaped
patterns in the turbulent states \cite{Brand}.

\subsection{The structure of the article}

\begin{figure}[tbp]
\begin{center}
\subfigure[]{\includegraphics[width=0.24\textwidth]{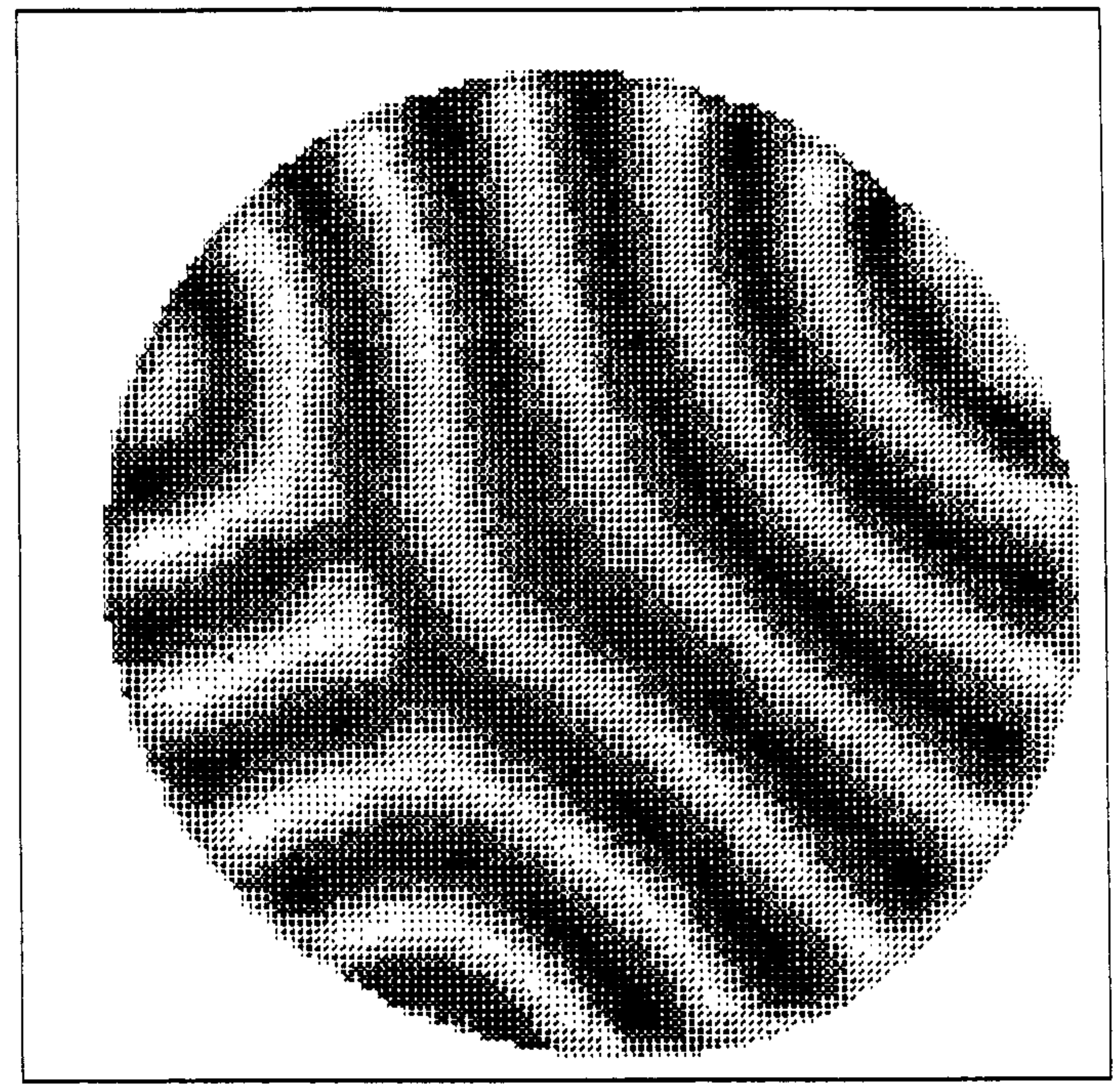}}%
\subfigure[]{\includegraphics[width=0.48\textwidth]{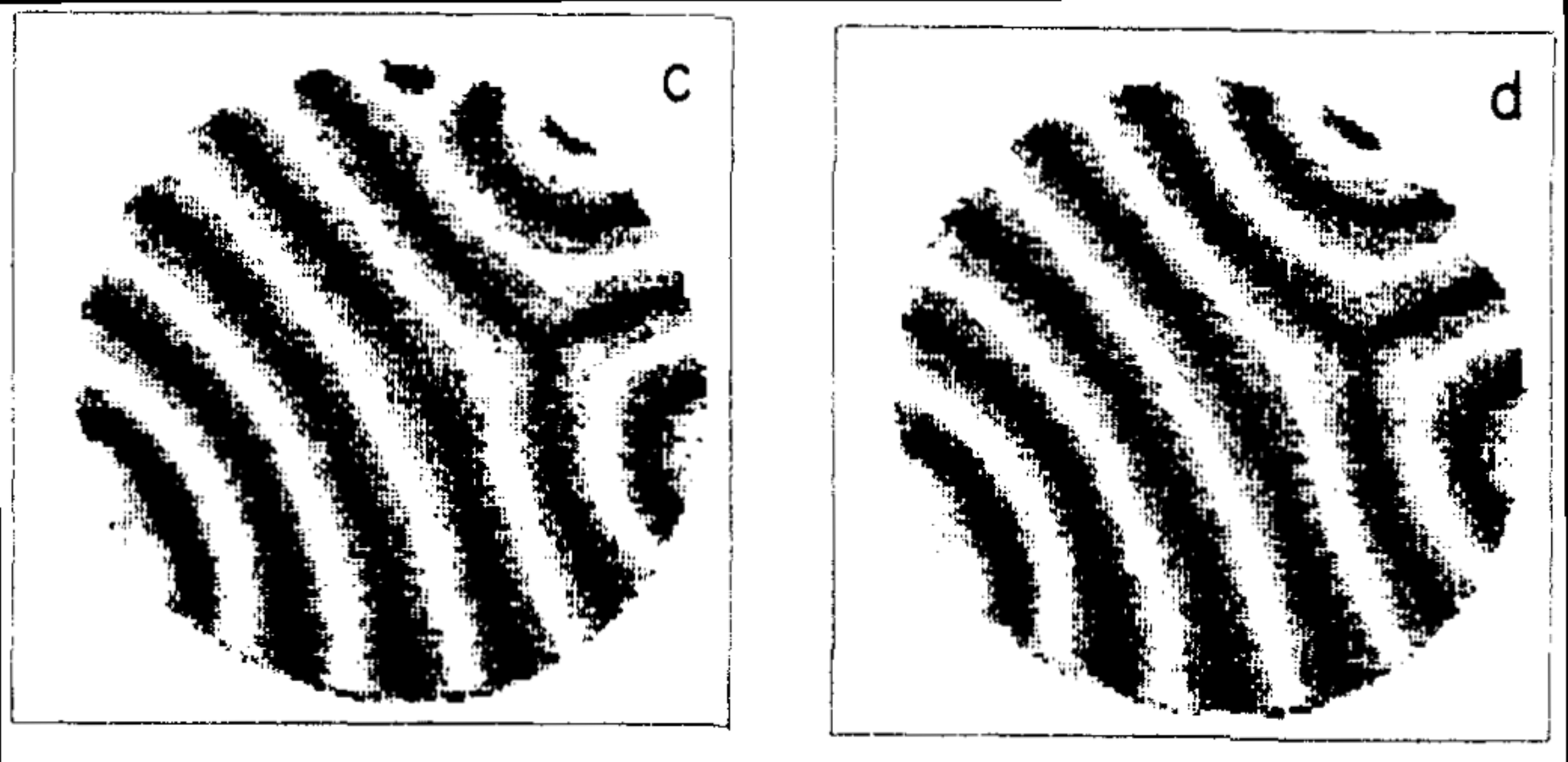}}
\end{center}
\caption{(a) An experimentally observed pattern of rolls in the Rayleigh-B%
\'{e}nard convection, which demonstrates a junction of domain walls (grain
boundaries). The pattern corresponds to overcriticality $\protect\varepsilon %
=1$, see Eq. (\protect\ref{varepsilon}). (b) Similar patterns observed at $%
\protect\varepsilon =1.8$ (left) and $2$ (right). Reprinted from Ref.
\protect\cite{Steinberg}.}
\label{fig0}
\end{figure}

The rest of the paper is divided into main Sections II and III. The former
one addresses the concept of DWs and its further development, following Ref.
\cite{Trib-DW}. The DWs considered in that work were constructed as
solutions of a system of two nonlinearly-coupled real GL equations, which
model the interaction of two families of simplest \textit{roll patterns}
(quasi-1D spatially periodic structures) in the Rayleigh-B\'{e}nard
convection. This setup is controlled by the overcriticality,%
\begin{equation}
\varepsilon =\left( \mathrm{Ra}-\mathrm{Ra}_{\mathrm{crit}}\right) /\mathrm{%
Ra}_{\mathrm{crit}}~,  \label{varepsilon}
\end{equation}%
where $\mathrm{Ra}$ is the Rayleigh number, and $\mathrm{Ra}_{\mathrm{crit}}$
is its critical value at the threshold of the instability of the fluid layer
heated from below. DWs in convection patterns were predicted as linear
defects (\textquotedblleft grain boundaries") \cite%
{Cross1982,Manneville,Iooss}, and were directly observed in experiments,
both as DWs proper and more complex structures, formed by junctions of DWs.
Typical examples of the experimentally observed patterns, borrowed from Ref.
\cite{Steinberg}, are presented in Fig. \ref{fig0}.

It is relevant to stress that the concept of grain boundaries is known, in a
great variety of different realizations, as a very general one in
condensed-matter physics \cite{grain1,grain2,grain4,grain5,grain6,grain3}.
In most cases, the nature of such objects is different from that in thermal
convection and other nonlinear dissipative media, Nevertheless, the
phenomenology of the grain boundaries in completely different physical
systems has many common features.

The DW states were constructed in Ref. \cite{Trib-DW} as solutions of two
coupled real GL equations for amplitudes of PWs connected by the DW. In a
particular case, such a solution for a symmetric DW is available in an exact
analytical form, see Eqs. (\ref{G=3}) and (\ref{exact}) below. It is also
demonstrated that the symmetric DW may play the role of a \textit{potential
well} which traps an additional small-amplitude component, in the form of a
bright soliton, thus making the structure of the DW more complex, as shown
below by Eqs. (\ref{exact-v})-(\ref{DD}) and Fig. \ref{fig_extra2}. Further,
a newly derived extension of the exact solution in included, for the case
when the symmetrically coupled real GL equations include linear-mixing terms
(see Eq. (\ref{exact2}) below), and a new exact solution for a strongly
asymmetric DW, in the case when only one real GL equation includes the
diffusion term (second derivative). This solution is given below by Eqs. (%
\ref{TF})-(\ref{xi}) and Fig. \ref{fig_extra1}.

At the level of stationary solutions, the same coupled equations which model
the grain boundaries in thermal convection predict DWs in optics, as
boundaries between spatial or temporal domains occupied by PWs representing
different polarizations or different carrier frequencies of light \cite%
{optical-DW}. These equations also produce DW states in binary Bose-Einstein
condensates (BECs) composed of immiscible components \cite{Poland}.

Still earlier, approximate solutions similar to DWs were constructed in the
framework of a single complex GL equation \cite{CGL-DW}. Such solutions
represent \textit{stationary sources} of stable PWs with wavenumbers $\pm k$
(see Eq. (\ref{PW})), emitted in opposite directions (while the
above-mentioned \textquotedblleft holes" \cite{holes} are \textit{sinks}
absorbing colliding PWs with opposite wavenumbers). A special case
corresponds to the complex GL equation (\ref{CCGL}) without the diffusion
term, i.e., with $a=0$. In that case, DWs may be approximately reduced to
shock waves governed by an effective Burgers equation for a local wavenumber
\cite{Burgers}. These results are also included, in a brief form, in Section
II. As an extension of the topic, this section also addresses DWs between
semi-infinite domains filled by counterpropagating traveling waves (this is
possible, in particular, in thermal convection in a layer of a binary fluid
heated from below \cite{Cross,Cross2,Coullet,Kolodner}). Furthermore,
Section II includes a newly found \textit{exact solution} for the DW between
traveling waves, produced by a system of coupled real GL equations that
include group-velocity terms (see Eqs. (\ref{r11})-(\ref{sgn}) below).

Section III summarizes some theoretical results for QP patterns in 2D and 3D
nonlinear dissipative media, the study of which was initiated in Ref. \cite%
{Trib-quasi}. In particular, included are findings for stable QP states
produced by combinations of four spatial modes in a laser cavity with
different 3D\ wave vectors \cite{Komarova}. Another possibility to produce a
spatially confined four-mode (eight-fold) QP structure, briefly mentioned in
Section III, is offered by the overlap of two square-shaped (two-mode)
patterns, under the angle of $45^{\text{o}}$, in a transient layer between
the patterns \cite{Rotstein}. This possibility is a combination of the two
main topics considered in this article, \textit{viz.}, DWs and QP patterns.

The article is completed by Section IV, which summarizes basic results and
briefly outlines new possibilities in this area.

\section{DW (domain-wall) patterns}

\subsection{The source pattern generated by the single complex GL equation}

\subsubsection{The generic case}

To produce approximate solutions to Eq. (\ref{CCGL}), it is convenient to
rewrite it in the Madelung form (\ref{Madelung}), which yields the following
system of equations for real amplitude $A$ and phase $\phi $:%
\begin{gather}
\frac{\partial A}{\partial t}=A-A^{3}+\frac{\partial ^{2}A}{\partial x^{2}}%
-A\left( \frac{\partial \phi }{\partial x}\right) ^{2}-2b\frac{\partial A}{%
\partial x}\frac{\partial \phi }{\partial x}-bA\frac{\partial ^{2}\phi }{%
\partial x^{2}},  \label{A} \\
A\frac{\partial \phi }{\partial t}=2\frac{\partial A}{\partial x}\frac{%
\partial \phi }{\partial x}+A\frac{\partial ^{2}\phi }{\partial x^{2}}%
-cA^{3}+b\frac{\partial ^{2}A}{\partial x^{2}}-bA\left( \frac{\partial \phi
}{\partial x}\right) ^{2}  \label{phi}
\end{gather}%
(recall coefficients of Eq. (\ref{CCGL}) are subject to normalization
conditions (\ref{11})). As shown in Ref. \cite{CGL-DW}, a stationary
solution of the DW type, which represents a source of PWs emitted in the
directions of $x\rightarrow \pm \infty $, can be looked for assuming that
the dispersion coefficients $b$ and $a$ are small, and the local amplitude, $%
A(x)$, and wavenumber, $p(x)\equiv \partial \phi /\partial x$, are slowly
varying functions of $x$ (the \textquotedblleft nonlinear geometric-optics
approximation", alias the \textquotedblleft \textit{eikonal approximation}%
"). In the lowest order, all derivatives and dispersion terms may be
neglected in Eq. (\ref{A}), reducing it merely to $A^{2}\approx 1-p^{2}$,
cf. Eq. (\ref{CW}). Next, this approximation is substituted in Eq. (\ref{phi}%
), with the phase taken as%
\begin{equation}
\phi (x,t)=-\left( c+(b-c)k^{2}\right) t+\int p(x)dx,  \label{phase}
\end{equation}%
where it is assumed that the asymptotic values of the wavenumber are
\begin{equation}
p\left( x\rightarrow \pm \infty \right) =\pm k  \label{pk}
\end{equation}%
(hence the frequency in expression (\ref{phase}) is the same as in Eq. (\ref%
{PW})). Keeping the lowest-order small terms with respect to the small
dispersive coefficients and small derivative $dp/dx$ of the slowly varying
local wavenumber leads to the following approximate equation:%
\begin{equation}
\frac{1-3p^{2}}{1-p^{2}}\frac{dp}{dx}=\left( c-b\right) \left(
k^{2}-p^{2}\right) .  \label{dp/dx}
\end{equation}

The DW solution to Eq. (\ref{dp/dx}) can be obtained in an implicit form,
which yields $x$ as a function of $p$, satisfying the boundary conditions (%
\ref{pk}):%
\begin{equation}
2k\ln \frac{1-p}{1+p}+\left( 1-3k^{2}\right) \ln \frac{k-p}{k+p}=2k\left(
b-c\right) \left( 1-k^{2}\right) x.  \label{implicit}
\end{equation}%
The solution can be easily cast in an explicit form under condition $%
k^{2}\ll 1$:%
\begin{equation}
p(x)\approx k\tanh \left[ \left( c-b\right) kx\right] .  \label{explicit}
\end{equation}%
This form clearly demonstrates that the DW may be indeed construed as an
emitter of waves from the center, where $p(x=0)=0$, to $x\rightarrow \pm
\infty $, in agreement with Eq. (\ref{dp/dx}). The explicit solutions, as
well as the implicit ones (\ref{implicit}), constitute a family
parameterized by free constant $k$.

In the real GL equation, with $b=c=0$, as well as in the case when the
linear and nonlinear dispersions exactly cancel each other, $b=c$, Eq. (\ref%
{dp/dx}) cannot produce a stationary DW solution. As shown in Ref. \cite%
{CGL-DW}, in that case initial configurations in the form resembling
expression (\ref{explicit}), i.e., a step-shaped profile of the local
wavenumber, give rise to nonstationary solutions, which may be approximated
by means of characteristics and caustics of a quasi-linear evolution
equation for $p(x,t)$.

\subsubsection{Domain walls as shock waves in the diffusion-free complex GL
equation}

The consideration of DWs should be performed differently in the special case
of the complex GL equation (\ref{CCGL}) with $a=0$, which does not include
the diffusion term. Taking into regard normalization (\ref{11}), the
respective equation takes the form of
\begin{equation}
\frac{\partial u}{\partial t}=u+ib\frac{\partial ^{2}u}{\partial x^{2}}%
-\left( 1+ic\right) |u|^{2}u  \label{a=0}
\end{equation}%
(in fact, one can additionally rescale coordinate $x$ here, to set $b=\pm 1$%
). This form of the equation admits free motion of various modes \cite%
{Burgers,Sakaguchi}. In this case, the Madelung substitution\ (\ref{Madelung}%
) leads, instead of the amplitude-phase equations (\ref{A}) and (\ref{phi}),
to ones%
\begin{gather}
\frac{\partial A}{\partial t}=A-A^{3}-2b\frac{\partial A}{\partial x}\frac{%
\partial \phi }{\partial x}-bA\frac{\partial ^{2}\phi }{\partial x^{2}},
\label{A2} \\
A\frac{\partial \phi }{\partial t}=-cA^{3}+b\frac{\partial ^{2}A}{\partial
x^{2}}-bA\left( \frac{\partial \phi }{\partial x}\right) ^{2}.  \label{phi2}
\end{gather}%
Further, the lowest approximation of the nonlinear geometric optics, applied
to Eq. (\ref{A2}), yields%
\begin{equation}
A^{2}\approx 1-b\frac{\partial p}{\partial x}.  \label{b}
\end{equation}%
The substitution of this in Eq. (\ref{phi2}) leads, after simple
manipulations (including the division by $A$ and differentiation with
respect to $x$, in order to replace $\partial \phi /\partial t$ by $\partial
p/\partial t$), to the Burgers equation \cite{Burgers},%
\begin{equation}
\frac{\partial p}{\partial t}=bc\frac{\partial ^{2}p}{\partial x^{2}}-2bp%
\frac{\partial p}{\partial x}.  \label{Burg}
\end{equation}%
The usual shock-wave solutions of Eq. (\ref{Burg}) give rise to a family of
DWs with two independent parameters, \textit{viz}., wall thickness $\xi >0$
and speed $s$, which may be positive or negative:%
\begin{equation}
p(x,t)=\frac{s}{2b}-\frac{c}{\xi }\tanh \left( \frac{x-st}{\xi }\right) .
\label{shock}
\end{equation}%
The appearance of the second free parameter, $s$, in this solution
corresponds to the above-mentioned fact that Eq. (\ref{a=0}) admits free
motion of patterns produced by this variant of the complex GL equation.

\subsection{DWs in systems of real coupled GL equations, as per Ref.
\protect\cite{Trib-DW}, and additional analytical findings}

\subsubsection{The setting}

The starting point of the analysis developed in Ref. \cite{Trib-DW} was a
general expression for the distribution of the complex order parameter in
the 2D system (e.g., the amplitude of the convective flow):%
\begin{equation}
U(x,y;t)=\sum_{l=1}^{N}u_{l}(x,y;t)\exp \left( i\mathbf{n}_{l}\mathbf{\cdot R%
}\right) ,  \label{U}
\end{equation}%
where $\mathbf{R=}\left( x,y\right) $. Equation (\ref{U}) implies that the
order-parameter field is a superposition of $N$ plane-wave modes (often
called \textit{rolls}, in the context of the convection theory) with wave
vectors $\mathbf{n}_{l}$, and $u_{l}\left( x,y\right) $ are slowly varying
amplitudes of these modes. Stationary states produced by the real GL
equations may be looked for in the real form too,
\begin{equation}
u_{l}(x,y)\equiv r_{l}\left( x,y\right) ,\arg \left( u_{l}\right) =0.
\label{r}
\end{equation}%
as the evolution of phases $\arg \left( u_{l}\right) $ is trivial in this
case.

It is relevant to mention that patterns similar to the rolls (known under
the same name) are produced by the Lugiato-Lefever (LL) equation and its
varieties. The basic LL equation may be considered as the NLS equation for
amplitude $u\left( x,t\right) $ of the optical field in a laser cavity,
which includes the linear-loss coefficient, $\gamma >0$, a real
cavity-mismatch parameter, $\theta \gtrless 0$, and a constant pump field, $%
u_{0}$ \cite{Lug-Lef}:%
\begin{equation}
\frac{\partial u}{\partial t}=-\gamma u+u_{0}+i\left( |u|^{2}-\theta \right)
u+i\frac{\partial ^{2}u}{\partial x^{2}}.  \label{u0}
\end{equation}%
Roll patterns were studied in details in various forms of LL models \cite%
{Oppo,Menyuk,Turing}. DWs also occur in these systems \cite{Staliunas,Oppo2}.

\begin{figure}[tbp]
\begin{center}
\subfigure[]{\includegraphics[width=0.40\textwidth]{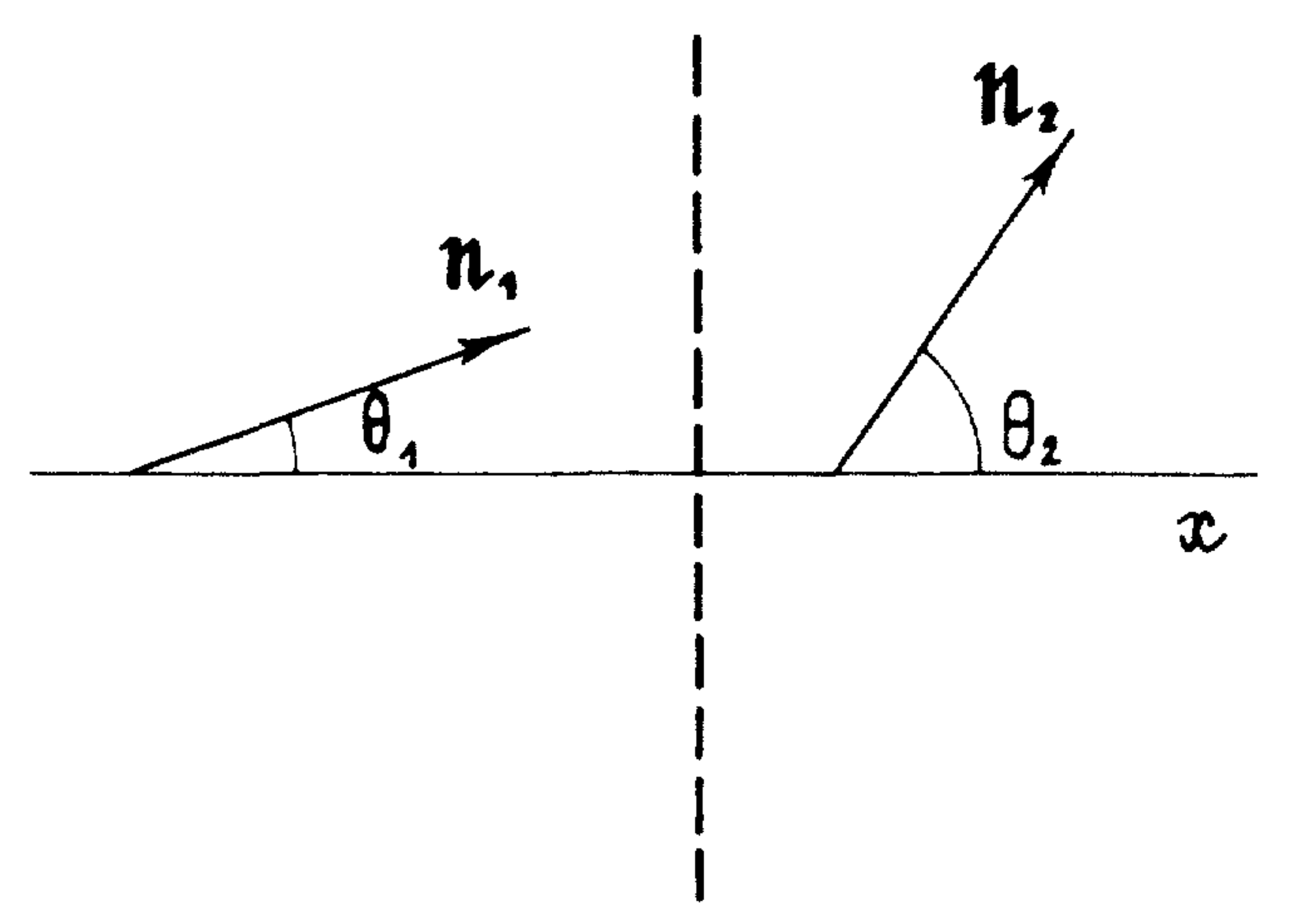}}%
\subfigure[]{\includegraphics[width=0.40\textwidth]{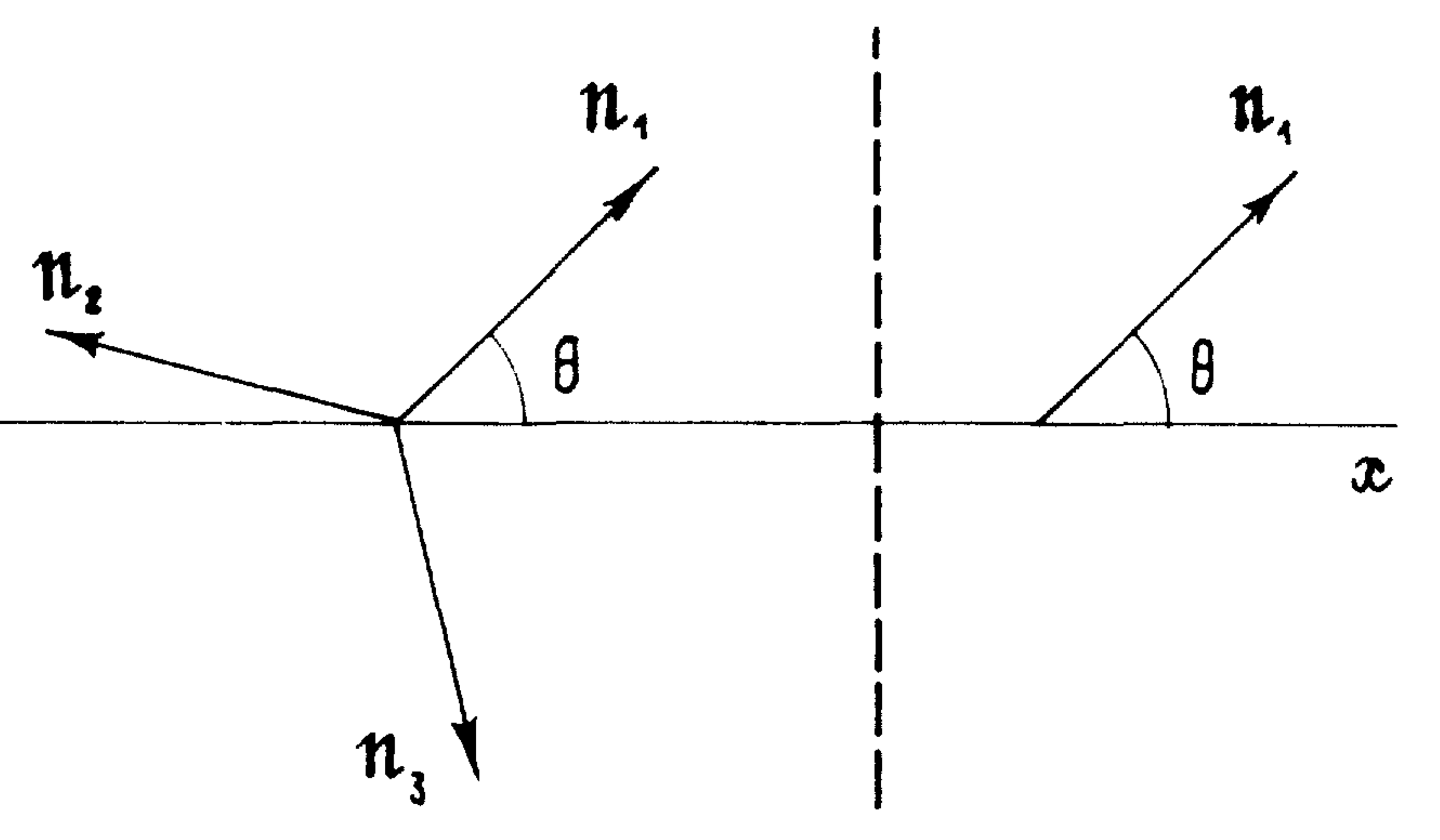}}
\end{center}
\caption{The scheme of the formation of the DW (domain wall)\ between 2D
patterns in the Rayleigh-B\'{e}nard convection and similar settings. (a) The
DW between plane waves (\textit{rolls}) with wave vectors oriented under
angles $\protect\theta _{1}$ and $\protect\theta _{2}$ with respect to the $%
x $ axis, see Eq. (\protect\ref{U}). The respective amplitudes $r_{1,2}(x)$
satisfy Eqs. (\protect\ref{r1}) and (\protect\ref{r2}) and are subject to
b.c. (\protect\ref{bc}). The position of the DW is shown by the vertical
dashed line. An example of the DW profile is displayed below in Fig. \protect
\ref{fig2}(a). (b) The same as in (a), for the DW between hexagons (the
triple-mode pattern) and single-mode rolls. The figure is reprinted from
Ref. \protect\cite{Trib-DW}.}
\label{fig1}
\end{figure}

As is illustrated by Fig. \ref{fig1}(a), the simplest possibility of the
realization of patterns represented by Eq. (\ref{U}) is the superposition of
$N=2$ modes, each one filling, essentially, a half-plane bounded by the DW.
In this case, real amplitudes $r_{1.2}$ are slowly varying functions of only
one coordinate, $x$, directed perpendicular to the DW. The respective
boundary conditions (b.c.) are%
\begin{gather}
r_{1}\left( x\rightarrow -\infty \right) =r_{2}\left( x\rightarrow +\infty
\right) =\mathrm{const}\neq 0,  \notag \\
r_{1}\left( x\rightarrow +\infty \right) =r_{2}\left( x\rightarrow -\infty
\right) =0.  \label{bc}
\end{gather}%
The scaled form of stationary (time-independent) coupled real GL equations
for slowly varying amplitudes $r_{1}(x)$ and $r_{2}(x)$, corresponding to
the bimodal DW configuration defined as per Fig. \ref{fig1}(a) and Eq. (\ref%
{bc}), are \cite{Trib-DW} (see also Refs. \cite{Cross1982} and \cite%
{Manneville})
\begin{eqnarray}
D_{1}\frac{d^{2}r_{1}}{dx^{2}}+r_{1}\left( 1-r_{1}^{2}-Gr_{2}^{2}\right)
&=&0,  \label{r1} \\
D_{2}\frac{d^{2}r_{2}}{dx^{2}}+r_{2}\left( 1-r_{2}^{2}-Gr_{1}^{2}\right)
&=&0.  \label{r2}
\end{eqnarray}%
Here effective diffusion coefficients are
\begin{equation}
D_{1.2}=\cos ^{2}\theta _{1,2}  \label{D}
\end{equation}%
(see Fig. \ref{fig1}(a)), and $G>0$ is an effective coefficient of the
cross-interaction between different plane waves, while the self-interaction
coefficient is scaled to be $1$.

The symmetric configuration corresponds to Fig. \ref{fig1}(a) with
\begin{equation}
\theta _{1}=-\theta _{2},  \label{symm}
\end{equation}%
which implies $D_{1}=D_{2}\equiv D$, according to Eq. (\ref{D}). Naturally,
the symmetric case plays an important role in the analysis, as shown below.

It is relevant to mention that coupled equations (\ref{r1}) and (\ref{r2})
may be considered as formal equations of motion for a mechanical system with
two degrees of freedom, while $x$ plays the role of formal time. This system
keeps a constant value of its (formal) Hamiltonian,%
\begin{equation}
h=\frac{1}{2}\sum_{j=1,2}\left[ D_{j}\left( \frac{dr_{j}}{dx}\right)
^{2}+r_{j}^{2}-\frac{1}{2}r_{j}^{4}\right] -\frac{G}{2}r_{1}^{2}r_{2}^{2}.
\label{h}
\end{equation}

\begin{figure}[tbp]
\begin{center}
\subfigure[]{\includegraphics[width=0.40\textwidth]{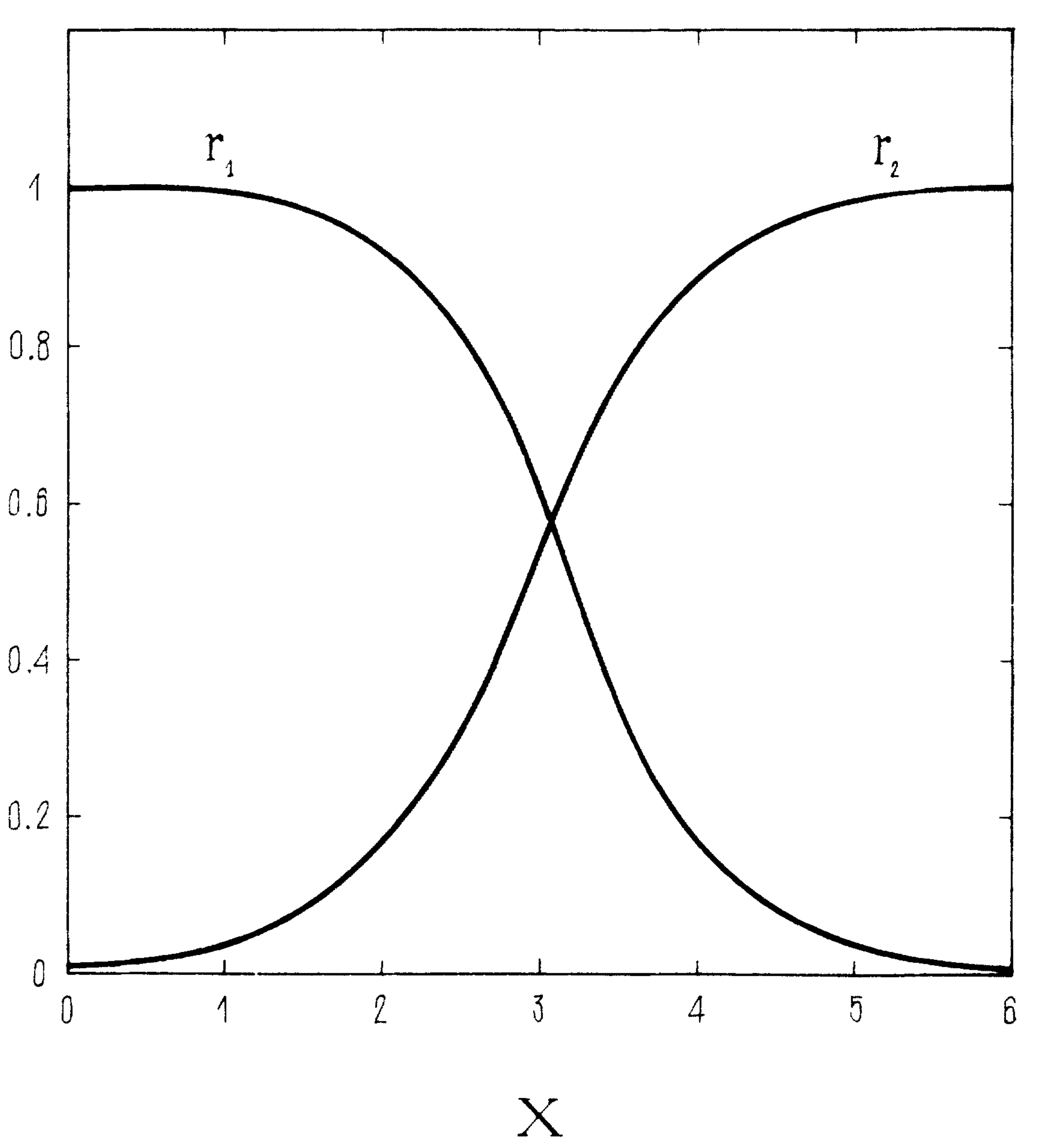}}%
\subfigure[]{\includegraphics[width=0.40\textwidth]{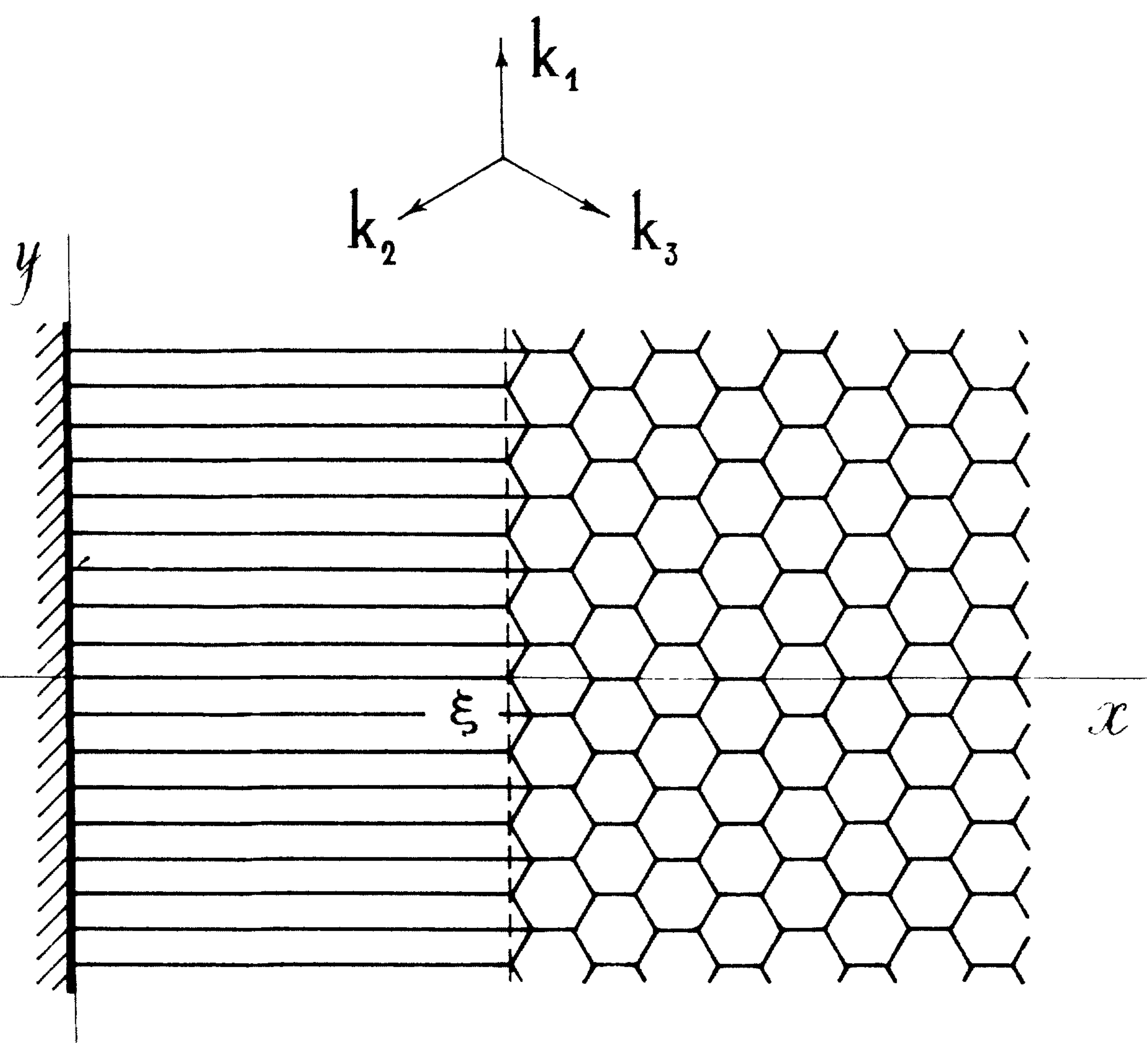}}
\end{center}
\caption{(a) A typical profile of the DW between different plane-wave (roll)
families. (b) The structure of the DW between the plane-wave and hexagonal
patterns (in this panel, $\mathbf{k}_{1,2,3}$ are identical to $\mathbf{n}%
_{l}$ in Fig. (\protect\ref{U})). The figure is reprinted from Ref.
\protect\cite{Trib-DW}.}
\label{fig2}
\end{figure}

DW solutions can be readily found as numerical solutions of coupled
equations (\ref{r1}) and (\ref{r2}), subject to b.c. (\ref{bc}). A
characteristic example of the solution is displayed in Fig. \ref{fig2}(a).
In fact, the existence of the DWs in the framework of Eqs. (\ref{r1}) and (%
\ref{r2}) may be understood as \textit{immiscibility} of the modes whose
amplitudes are produced by these equations. The general condition for the
immiscibility, in the present notation, is well known since long ago:
\begin{equation}
G>1,  \label{G>1}
\end{equation}
i.e., the strength of the mutual repulsion of the two components must exceed
the strength\ of their self-repulsion \cite{Mineev}.

Hexagonal states in the Rayleigh-B\'{e}nard convection are produced by a
superposition of three plane waves, with angles $120^{\mathrm{o}}$ between
their wave vectors. Such patterns are stable if, in addition to the cubic
inter-mode interaction in Eqs. (\ref{r1}) and (\ref{r2}), the respective
system of three GL equations for local amplitudes $r_{1,2,3}(x)$ includes
resonant quadratic terms:%
\begin{equation}
D_{1}\frac{d^{2}r_{1}}{dx^{2}}+r_{1}\left[ 1-r_{1}^{2}-G\left(
r_{2}^{2}+r_{3}^{2}\right) \right] +\nu r_{2}r_{3}=0  \label{triple}
\end{equation}%
plus two complementing equations obtained from Eq. (\ref{triple}) by cyclic
permutations of subscripts $\left( 1,2,3\right) $, where $\nu $ is a
coefficient of the resonant interaction. In the theory of the thermal
convection, the quadratic terms represent effects beyond the framework of
the basic Boussinesq approximation \cite{Busse,Pomeau}. Numerical solution
of Eq. (\ref{triple}) produces DWs connecting single-mode rolls and the
hexagonal pattern \cite{Trib-DW}, see an illustration in Fig. \ref{fig1}(b)
and the corresponding pattern displayed in Fig. \ref{fig2}(b). It was also
demonstrated that DWs are possible between two bimodal (square-shaped)
patterns, each one composed of two plane waves with perpendicular
orientations. In that case, the DW appears as a boundary between two
half-planes filled by square patterns with different orientations \cite%
{Trib-DW,Rotstein}, see further details below in Eqs. (\ref{34})-(\ref{B2})
and Fig. \ref{fig6}. In Ref. \cite{Martin}, a spatially inhomogeneous model,
in which the cross-interaction coefficient is a function of the coordinate, $%
G=G(x)$, was introduced, making it possible to construct stable DWs between
the single- and bimodal patterns.

\subsubsection{Basic analytical results from Ref. \protect\cite{Trib-DW}}

An analytically tractable case is the symmetric one, with $D_{1}=D_{2}\equiv
D$, and
\begin{equation}
0<G-1\ll 1  \label{G-1}
\end{equation}%
(recall that $G>1$ is a necessary condition for the existence of DWs). The
analysis makes it possible to reduce these coupled equations to an effective
sine-Gordon equation for a slowly varying inter-component phase $\chi (x)$,
and thus produce an approximate analytical DW solution with a large width of
the transient layer, $L\sim \left( G-1\right) ^{-1/2}$ \cite{Trib-DW}:%
\begin{equation}
\left\{
\begin{array}{c}
r_{1}(x) \\
r_{2}(x)%
\end{array}%
\right\} \approx \left\{
\begin{array}{c}
\cos \chi (x) \\
\sin \chi (x)%
\end{array}%
\right\} ,\chi =\arctan \left( \exp \left( \sqrt{\frac{G-1}{D}}x\right)
\right) .  \label{chi}
\end{equation}

In the particular case of
\begin{equation}
G=3,  \label{G=3}
\end{equation}%
the symmetric version of Eqs. (\ref{r1}) and (\ref{r2}) admits an \emph{exact%
} DW solution \cite{Trib-DW}:%
\begin{equation}
\left\{
\begin{array}{c}
r_{1}(x) \\
r_{2}(x)%
\end{array}%
\right\} =\frac{1}{2}\left\{
\begin{array}{c}
1-\tanh \left( x/\sqrt{2D}\right) \\
1+\tanh \left( x/\sqrt{2D}\right)%
\end{array}%
\right\} ,  \label{exact}
\end{equation}%
which is obviously compatible with b.c. (\ref{bc}).

\subsubsection{New analytical results}

\paragraph{Symmetric DWs}

Precisely the same real time-independent equations as Eqs. (\ref{r1}) and (%
\ref{r2}) appear in nonlinear optics as a stationary version of coupled NLS
equations in bimodal waveguides, with $r_{1}$ and $r_{2}$ being local
amplitudes of electromagnetic waves carrying different wavelengths or
different polarizations of light \cite{KA}. In the latter case, typical
values of $G$ are $2/3$ or $2$, for the linear or circular polarizations of
the light, respectively. Other values are possible too, in photonic-crystal
waveguides \cite{phot-cryst}. Similarly, the stationary real equations
naturally appear as the time-independent version of coupled Gross-Pitaevskii
equations for mean-field wave functions of binary BECs in ultracold atomic
gases \cite{Pit}.

Thus, the same solutions as considered here may represent optical DWs \cite%
{optical-DW}, as well as DWs separating two immiscible species in the BEC
\cite{Poland}. Further, the coupled equations in optics and BEC\ models may
also include linear mixing between the interacting modes. In particular,
this effect is produced by twist applied to the bulk optical waveguide. A
similar effect in binary BEC, \textit{viz}., mutual inter-conversion of two
atomic states, which form the binary BEC, may be induced by the resonant
radio-frequency field \cite{radio}. The respectively modified symmetric
system of Eqs. (\ref{r1}) and (\ref{r2}) is
\begin{eqnarray}
D\frac{d^{2}r_{1}}{dx^{2}}+r_{1}\left( 1-r_{1}^{2}-Gr_{2}^{2}\right)
+\lambda r_{2} &=&0,  \label{lambda1} \\
D\frac{d^{2}r_{2}}{dx^{2}}+r_{2}\left( 1-r_{2}^{2}-Gr_{1}^{2}\right)
+\lambda r_{1} &=&0,  \label{lambda2}
\end{eqnarray}%
where real $\lambda $ is the linear-coupling coefficient. In fact, Eqs. (\ref%
{lambda1}) and (\ref{lambda2}) apply to the Rayleigh-B\'{e}nard convection
too, in the case when periodic corrugation of the bottom of the convection
cell, with amplitude $\sim \lambda $ and wave vector $\mathbf{n}_{1}+\mathbf{%
n}_{2}$ (see Eq. (\ref{U})), gives rise to the effect of the \textit{linear
cross-gain}. Actually, it is used in many laser setups that are similar to
thermal convection \cite{cross-,cross-2}.

The system of Eqs. (\ref{lambda1}) and (\ref{lambda2}) with $G=3$ admits an
exact DW solution, which is an extension of its counterpart (\ref{exact}):%
\begin{equation}
\left\{
\begin{array}{c}
r_{1}(x) \\
r_{2}(x)%
\end{array}%
\right\} =\frac{1}{2}\left\{
\begin{array}{c}
\sqrt{1+\lambda }-\sqrt{1-\lambda }\tanh \left( \sqrt{\frac{1-\lambda }{2D}}%
x\right) \\
\sqrt{1+\lambda }+\sqrt{1-\lambda }\tanh \left( \sqrt{\frac{1-\lambda }{2D}}%
x\right)%
\end{array}%
\right\} .  \label{exact2}
\end{equation}%
Due to the action of the linear mixing, b.c. (\ref{bc}) are replaced by%
\begin{gather}
r_{1}\left( x\rightarrow -\infty \right) =r_{2}\left( x\rightarrow +\infty
\right) =\frac{1}{2}\left( \sqrt{1+\lambda }+\sqrt{1-\lambda }\right) ,
\notag \\
r_{1}\left( x\rightarrow +\infty \right) =r_{2}\left( x\rightarrow -\infty
\right) =\frac{1}{2}\left( \sqrt{1+\lambda }-\sqrt{1-\lambda }\right) .
\label{bc2}
\end{gather}%
The extended exact solution given by Eqs. (\ref{lambda1})-(\ref{bc2}) was
not reported in previous publications.

\paragraph{The effect of the confining potential}

The above-mentioned realization of the coupled real GL equations in terms of
the binary BEC should include, in the general case, a trapping
harmonic-oscillator (HO) potential, which is normally used in the experiment
\cite{Pit}. The accordingly modified system of Eqs. (\ref{lambda1}) and (\ref%
{lambda2}) is%
\begin{eqnarray}
D\frac{d^{2}r_{1}}{dx^{2}}+r_{1}\left( 1-r_{1}^{2}-Gr_{2}^{2}\right)
+\lambda r_{2} &=&\frac{\aleph ^{2}}{2}x^{2}r_{1},  \label{omega1} \\
D\frac{d^{2}r_{2}}{dx^{2}}+r_{2}\left( 1-r_{2}^{2}-Gr_{1}^{2}\right)
+\lambda r_{1} &=&\frac{\aleph ^{2}}{2}x^{2}r_{2},  \label{omega2}
\end{eqnarray}%
where $\aleph ^{2}$ is the strength of the OH potential. DW solutions of the
system of Eqs. (\ref{omega1}) and (\ref{omega2}) were addressed in Ref. \cite%
{Merh}. A rigorous mathematical framework for the analysis of such solutions
in the absence of the linear coupling ($\lambda =0$) was elaborated in Ref.
\cite{Peli}.

If the HO\ trap is weak enough, \textit{viz}., $\aleph ^{2}\ll 4/\left(
1-\lambda \right) $, the DW trapped in the OH potential takes nearly
constant values, close to those in Eq. (\ref{bc2}), in the region of%
\begin{equation}
2D/\left( 1-\lambda \right) \ll x^{2}\ll 8D/\aleph ^{2}.  \label{intermed}
\end{equation}%
At $x^{2}\rightarrow \infty $, solutions generated by Eqs. (\ref{omega1})
and (\ref{omega2}) decay similar to eigenfunctions of the HO potential in
quantum mechanics, \textit{viz}.,
\begin{eqnarray}
r_{1,2} &\approx &\varrho _{1,2}|x|^{\beta }\exp \left( -\frac{\aleph }{2%
\sqrt{2D}}x^{2}\right) ,  \label{r12} \\
\beta &=&\frac{1+\lambda }{\sqrt{2D}\aleph }-\frac{1}{2},  \label{beta}
\end{eqnarray}%
where $\varrho _{1,2}$ are constants. In the case of $\lambda =0$, the
asymptotic tails (\ref{r12}) follow the structure of solution (\ref{exact}),
i.e., $\varrho _{1}\left( x\rightarrow +\infty \right) =\varrho _{2}\left(
x\rightarrow -\infty \right) =0$ and $\varrho _{1}\left( x\rightarrow
-\infty \right) =\varrho _{2}\left( x\rightarrow +\infty \right) \neq 0$. On
the other hand, the presence of the linear mixing, $\lambda \neq 0$, makes
the tail symmetric with respect to the two components, $\varrho _{1}\left(
|x|\rightarrow \infty \right) =\varrho _{2}\left( |x|\rightarrow \infty
\right) \neq 0$. Note that $\beta =0$ in Eq. (\ref{beta}) with $\lambda =0$
is tantamount to the case when values of $\aleph $ and $D$ in Eqs. (\ref%
{omega1}) and (\ref{omega2}) correspond to the ground state of the HO
potential.

\paragraph{Exact asymmetric DWs}

Another possibility to add a new analytical solution for DWs appears in the
limit case of the extreme asymmetry in the system of Eqs. (\ref{r1}) and (%
\ref{r2}), which corresponds to $D_{2}=0$ and $D_{1}\equiv D>0$, i.e., the
DW between two roll families one of which has the wave vector perpendicular
to the $x$ axis, see Eq. (\ref{D}):
\begin{gather}
D\frac{d^{2}r_{1}}{dx^{2}}+r_{1}\left( 1-r_{1}^{2}-Gr_{2}^{2}\right) =0,
\label{D1} \\
r_{2}\left( 1-r_{2}^{2}-Gr_{1}^{2}\right) =0.  \label{D=0}
\end{gather}%
Actually, the form of Eq. (\ref{D=0}), in which the second derivative drops
out, corresponds to the well-known Thomas-Fermi approximation (TF)\ in the
BEC\ theory. In the framework of the TF approximation, the kinetic-energy
term in the Gross-Pitaevskii equation is neglected, in comparison with
larger ones, representing a trapping potential and the (self-repulsive)
nonlinearity \cite{Pit}. In the present case, corresponding to $\theta
_{2}=90^{\mathrm{o}}$, i.e., $D_{2}=0$ in Eq. (\ref{D}), is not an
approximation, but the exact special case. As concerns the application of
Eqs. (\ref{r1}) and (\ref{r2}) to BEC, with the kinetic-energy coefficients $%
D_{1,2}=\hbar ^{2}/\left( 2m_{1,2}\right) $ in physical units, where $%
m_{1,2} $ are atomic masses of the two components of the \textit{%
heteronuclear} binary condensate, Eqs. (\ref{D1}) and (\ref{D=0}) correspond
to a \textit{semi-TF approximation}, representing a mixture of light (small $%
m_{1}$) and heavy (large $m_{2}$) atoms, e.g., a $^{7}$Li--$^{87}$Rb
diatomic gas \cite{Li-Rb}.

Obviously, Eq. (\ref{D=0}) yields two solutions, \textit{viz}., either%
\begin{equation}
r_{2}=0,  \label{r2=0}
\end{equation}%
or one featuring to the quasi-TF relation,
\begin{equation}
r_{2}^{2}(x)=1-Gr_{1}^{2}(x).  \label{TF}
\end{equation}%
In the former case, Eq. (\ref{D1}) with $r_{2}=0$ yields a solution in the
form of the usual dark soliton, while in the latter case, the substitution
of expression (\ref{TF}) in Eq. (\ref{D1}) produces a bright-soliton
solution. These solutions may be \textquotedblleft dovetailed" at a stitch
point,
\begin{equation}
x=x_{0}\equiv -\sqrt{2D}\ln \left( \frac{\sqrt{G}+1}{\sqrt{G}-1}\right) ,
\label{x0}
\end{equation}%
which is defined by condition $r_{1}^{2}(x)=1/G$ (see Eq. (\ref{TF}). The
global form of the solution, which complies with b.c. (\ref{bc}), is%
\begin{equation}
r_{1}(x)=\left\{
\begin{array}{c}
-\tanh \left( x/\sqrt{2D}\right) ,~\mathrm{at}~-\infty <x<x_{0}, \\
\sqrt{\frac{2}{G+1}}\mathrm{sech}\left[ \sqrt{\frac{G-1}{D}}\left( x-\xi
\right) \right] ,\mathrm{at}~x_{0}<x<+\infty ,%
\end{array}%
\right.  \label{r1exact}
\end{equation}%
\begin{equation}
r_{2}(x)=\left\{
\begin{array}{c}
0,~\mathrm{at}~-\infty <x<x_{0}, \\
\sqrt{1-Gr_{1}^{2}(x)},\mathrm{at}~x_{0}<x<+\infty .%
\end{array}%
\right.  \label{r2exact}
\end{equation}%
Finally, the virtual center of bright-soliton segment of $r_{1}(x)$ is
located at
\begin{equation}
x=\xi \equiv x_{0}-\sqrt{\frac{D}{G-1}}\ln \left( \sqrt{\frac{2G}{G+1}}+%
\sqrt{\frac{G-1}{G+1}}\right)  \label{xi}
\end{equation}%
(actually, exact solution (\ref{r1exact}) makes use of the \textquotedblleft
tail" of the bright soliton in the region of $x\geq x_{0}$, which does not
include the central point, $x=\xi $). The distance $x_{0}-\xi $, determined
by Eq. (\ref{xi}), defines the effective width of the strongly asymmetric
DW. Note that, as seen in Eqs. (\ref{x0}) and (\ref{r1exact}), this exact
solution exists under the condition of $G>1$, which is the above-mentioned
immiscibility condition.

\begin{figure}[tbp]
\begin{center}
\includegraphics[width=0.60\textwidth]{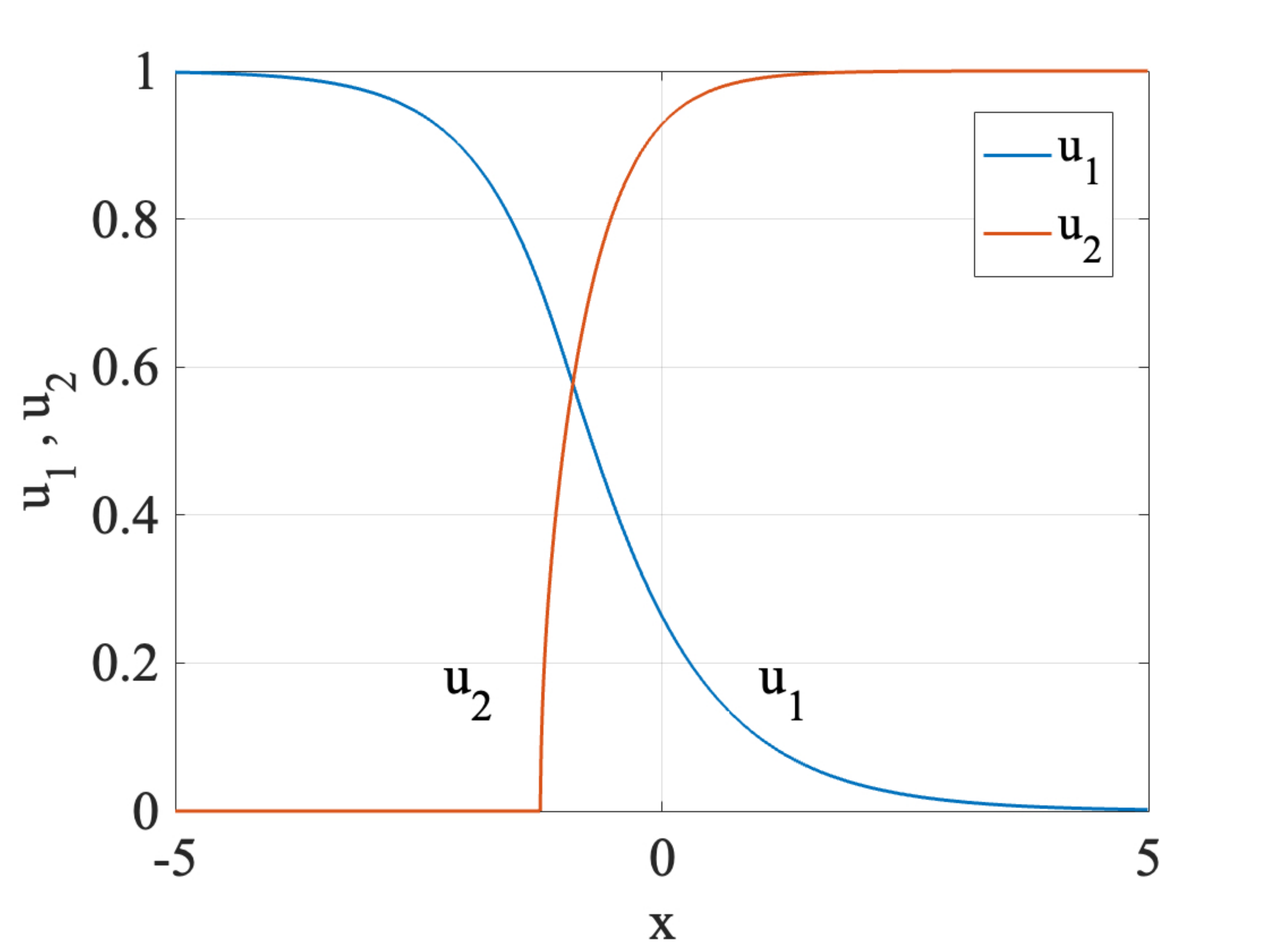}
\end{center}
\caption{An example of the asymmetric DW, as given by Eqs. (\protect\ref{x0}%
)-(\protect\ref{xi}), for $D=1$ and $G=2$ (here, $u_{1}$ and $u_{2}$ stand
for $r_{1}$ and $r_{2}$ in the analytical solution). Note that the
coordinate of the stitch point is, in this case, $x_{0}\approx -1.25$, as
per Eq. (\protect\ref{x0}), and the \textquotedblleft virtual center" of the
bright-soliton segment of $u_{1}(x)$ is located at $\protect\xi \approx
-1.80 $, as per Eq. (\protect\ref{xi}).}
\label{fig_extra1}
\end{figure}

It is easy to check that expression (\ref{r1exact}) satisfies the continuity
conditions for $r_{1}(x)$ and $dr_{1}/dx$ at $x=x_{0}$, and expression (\ref%
{r2exact}) provides the continuity of $r_{2}(x)$ at the same point. The
continuity of $dr_{2}/dt$ at $x=x_{0}$ is not required, as Eq. (\ref{D=0})
does not include derivatives. A typical example of the exact solution is
displayed, for $D=1$ and $G=2$, in Fig. \ref{fig_extra1}.

A remarkable fact is that, unlike the above-mentioned exact solutions (\ref%
{exact}) and (\ref{exact2}), which exist solely at $G=3$ (see Eq. (\ref{G=3}%
)), the one given by Eqs. (\ref{x0})-(\ref{xi}) exists, as a \textit{generic
one}, for all values of $G>1$.

\subsubsection{DW-bright-soliton complexes}

\paragraph{An exact solution for the composite state}

The DW formed by two immiscible PWs may serve as an effective potential for
trapping an additional PW mode. To address this possibility, it is relevant
to consider the symmetric configuration, with $D_{1}=D_{2}\equiv D$ (see Eq.
(\ref{D})), and wave vector $\mathbf{k}_{v}$ of the additional PW mode, $%
v(x) $, directed along the bisectrix of the angle between the DW-forming
wave vectors $\mathbf{k}_{1}$ and $\mathbf{k}_{2}$, i.e., along axis $x$
(hence Eq. (\ref{D}) yields $D_{v}=1$). The corresponding system of three
coupled stationary real GL equations is%
\begin{eqnarray}
D\frac{d^{2}u_{1}}{dx^{2}}+u_{1}\left( 1-u_{1}^{2}-Gu_{2}^{2}-gv^{2}\right)
&=&0,  \label{u1g} \\
D\frac{d^{2}u_{2}}{dx^{2}}+u_{2}\left( 1-u_{2}^{2}-Gu_{1}^{2}-gv^{2}\right)
&=&0,  \label{u2g}
\end{eqnarray}%
\begin{equation}
\frac{d^{2}v}{dx^{2}}+\left( 1-v^{3}-g\left( u_{1}^{2}+u_{2}^{2}\right)
\right) v=0,  \label{vv}
\end{equation}%
where $g>0$ is the constant of the nonlinear interaction between components $%
u_{1,2}$ and $v$.

The system of Eqs. (\ref{u1g})-(\ref{vv}) admits the following exact
solution, in the form of the DW of components $u_{1,2}(x)$ coupled to a
bright-soliton profile of $v(x)$:%
\begin{equation}
\left\{
\begin{array}{c}
u_{1}(x) \\
u_{2}(x)%
\end{array}%
\right\} =\frac{1}{2}\left\{
\begin{array}{c}
1-\tanh \left( \sqrt{g-1}x\right) \\
1+\tanh \left( \sqrt{g-1}x\right)%
\end{array}%
\right\} ,  \label{exact-v}
\end{equation}%
\begin{equation}
v(x)=\sqrt{2-\frac{3}{2}g}\mathrm{sech}\left( \sqrt{g-1}x\right) .
\label{exact-vv}
\end{equation}%
This solution is valid under the condition that coefficients $G$ and $D$ in
Eqs. (\ref{u1g}) and (\ref{u2g}) take the following particular values,%
\begin{equation}
G=3-8g+6g^{2},  \label{G}
\end{equation}%
\begin{equation}
D=\frac{1}{2}\left( 3g-1\right) .  \label{DD}
\end{equation}%
As is follows from Eq. (\ref{exact-vv}), this solution contains free
parameter $g$, which may vary in a narrow interval,
\begin{equation}
1<g<4/3  \label{4/3}
\end{equation}%
(see also Eq. (\ref{1-2}) below). According to Eqs. (\ref{G}) and (\ref{DD}%
), the interval (\ref{4/3}) corresponds to coefficients $G$ and $D$\ varying
in intervals%
\begin{equation}
1<G<3;~1<D<3/2.  \label{GD}
\end{equation}%
Thus, adding the $v$ component lifts the degeneracy of the exact DW solution
(\ref{exact}), which exists solely at $G=3$.

Recall that, in the model of convection patterns, $D$ cannot take values $%
D>1 $, which disagrees with Eq. (\ref{GD}). However, values $D>1$ are
relevant for systems of Gross-Pitaevskii equations for the heteronuclear
three-component BEC. In the latter case, $D$ is the ratio of atomic masses
of the different species which form the triple immiscible BEC. Similarly, $D$
is the ratio of values of the normal group-velocity dispersion of
copropagating waves in the temporal-domain realization of the real GL
equations in nonlinear fiber optics \cite{optical-DW}. In the latter case,
values $D>1$ are relevant too.

An example of the DW-bright-soliton complex is displayed in Fig. \ref%
{fig_extra2} for $g=7/6$, in which case Eqs. (\ref{G}) and (\ref{DD}) yield $%
G=11/6$ and $D=5/4$ (according to Eqs. (\ref{G}) and (\ref{D})).
\begin{figure}[tbp]
\begin{center}
\includegraphics[width=0.60\textwidth]{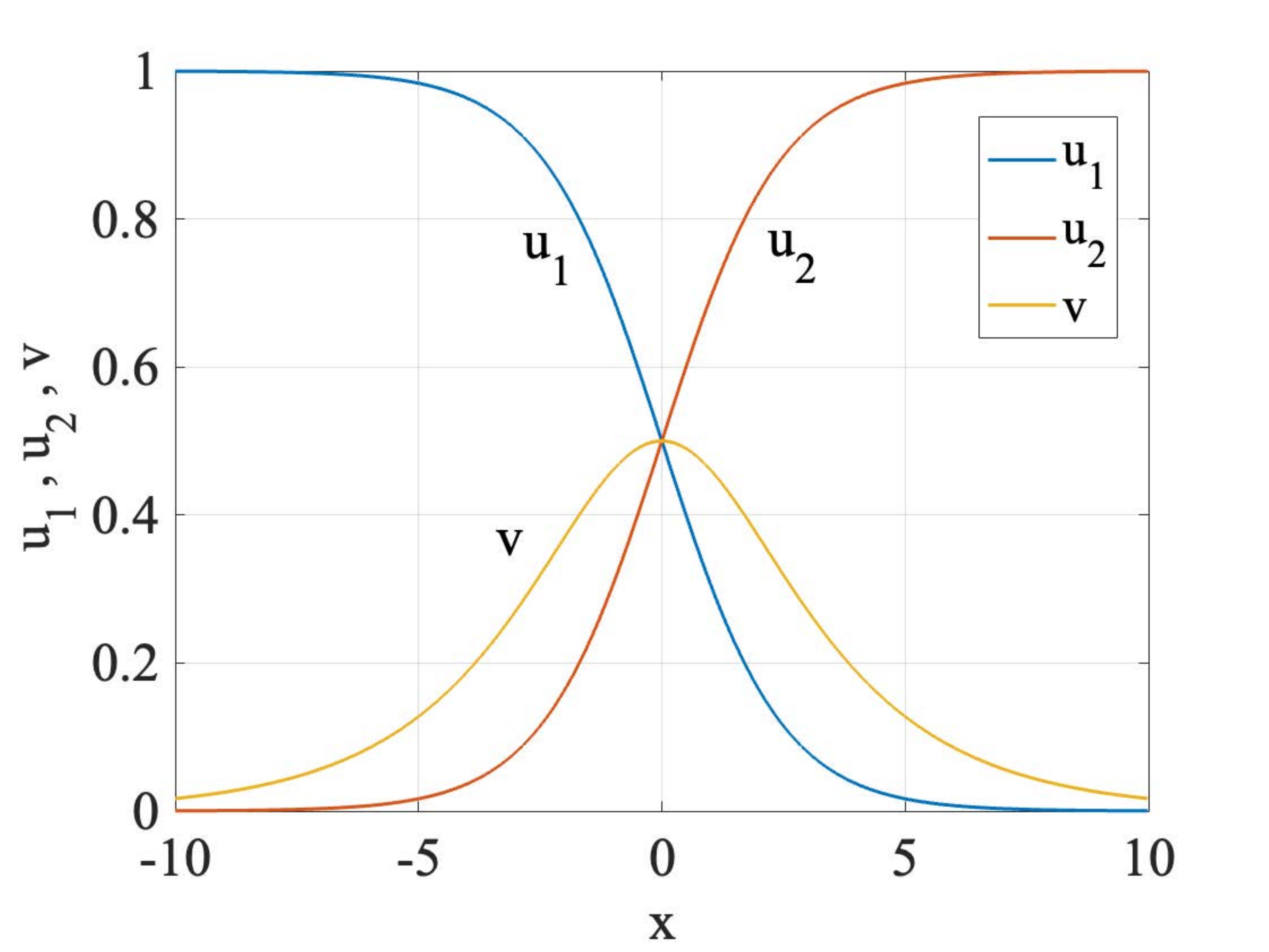}
\end{center}
\caption{An example of the exact solution for the DW-bright-soliton complex,
given by Eqs. (\protect\ref{exact-v}) and (\protect\ref{exact-vv}), for $%
g=7/6,G=11/6,$ and $D=5/4$.}
\label{fig_extra2}
\end{figure}

\paragraph{The bifurcation of the creation of the composite state in the
general case}

If relation (\ref{G}) is not imposed on the interaction coefficients $g$ and
$G$, the solution for the composite state cannot be found in an exact form.
Nevertheless, it is possible to identify \textit{bifurcation points} at
which component $v$ with an infinitesimal amplitude appears. To this end,
Eq. (\ref{vv}) should be used in the form linearized with respect to $v$:
\begin{equation}
\frac{d^{2}v}{dx^{2}}+\left\{ 1-g\left[ u_{1}^{2}(x)+u_{2}^{2}(x)\right]
\right\} v=0.  \label{rho}
\end{equation}%
This linear equation can be exactly solved for $u_{1}(x)=u_{2}(x)$ given by
expression (\ref{exact}), in the case of $G=3$, while parameters $D$ and $g$
may take arbitrary values. Indeed, using the commonly known solution for the
P\"{o}schl-Teller potential in quantum mechanics, it is easy to find that
Eq. (\ref{rho}) with the effective potential corresponding to expression
(\ref{exact}) gives rise to eigenmodes in the form of%
\begin{equation}
v(x)=\mathrm{const}\cdot \left[ \mathrm{sech}\left( x/\sqrt{2D}\right) %
\right] ^{\alpha },  \label{eigen}
\end{equation}%
at a special value of the interaction coefficient, which identifies the
bifurcation producing the composite state:%
\begin{equation}
g_{\mathrm{bif}}=D^{-1}\left( 1+2D\mp \sqrt{1+2D}\right) ,  \label{bif}
\end{equation}%
the respective value of power $\alpha $ in expression (\ref{eigen}) being%
\begin{equation}
\alpha =\sqrt{2\left( 1+D\mp \sqrt{1+2D}\right) }.  \label{alpha}
\end{equation}%
The values given by Eqs. (\ref{bif}) and (\ref{alpha}) with the top sign
from $\mp $ correspond to the bifurcation creating a fundamental composite
state (the ground state, in terms of the quantum-mechanical analog) at $g>g_{%
\mathrm{bif}}$, while the bottom sign represents a higher-order bifurcation
(alias the second excited state, in the language of quantum mechanics; the
first excited state, which is not considered here, is a spatially odd mode).
While it is obvious that the fundamental bifurcation provides a stable
composite state, it is plausible that the ones produced by higher-order
bifurcations are unstable.

Lastly, varying coefficient $D$ of the modes forming the underlying DW
between $D=0$ and $D=\infty $ (recall that the convection model corresponds
to $D<1,$ while the realizations in BEC and optics admit $D>1$), Eq. (\ref%
{bif}) demonstrates monotonous variation of the bifurcation point in
interval
\begin{equation}
g_{\mathrm{bif}}(D=0)\equiv 1<g_{\mathrm{bif}}<2\equiv g_{\mathrm{bif}%
}\left( D\rightarrow \infty \right) .  \label{1-2}
\end{equation}%
It extends interval (\ref{4/3}) in which exact composite states with a
finite amplitude were found above, see Eqs. (\ref{exact-v})-(\ref{DD}).

\subsection{Domain walls between traveling waves}

\subsubsection{The setting}

An essential extension of the above results for DWs, produced by coupled
equations (\ref{r1}) and (\ref{r2}), was reported in Ref. \cite{traveling},
which addressed a system of coupled GL equations for counter-propagating
traveling waves, such as those occurring in binary-fluid convection \cite%
{Cross,Cross2,Coullet,Kolodner}. The system is composed of two equations of
type (\ref{CCGL}) (subject to normalization (\ref{11})), coupled by complex
cubic terms with coefficients $G$ and $H$:
\begin{eqnarray}
\frac{\partial u_{1}}{\partial t}+s\frac{\partial u_{1}}{\partial x}
&=&u_{1}+\left( 1+ib\right) \frac{\partial ^{2}u_{1}}{\partial x^{2}}-\left(
1+ic\right) |u_{1}|^{2}u_{1}-\left( G+iH\right) |u_{2}|^{2}u_{1},  \label{u1}
\\
\frac{\partial u_{2}}{\partial t}-s\frac{\partial u_{1}}{\partial x}
&=&u_{2}+\left( 1+ib\right) \frac{\partial ^{2}u_{2}}{\partial x^{2}}-\left(
1+ic\right) |u_{2}|^{2}u_{2}-\left( G+iH\right) |u_{1}|^{2}u_{2},  \label{u2}
\end{eqnarray}%
where $-s$ and $+s$ are group velocities of the counter-propagating waves, $%
u_{1}$ and $u_{2}$ (real coefficient $H$ represents the \textit{cross-phase
modulation} (XPM), in terms of optics \cite{KA}). A natural approach to
constructing DW solutions of the system of Eqs. (\ref{u1}) and (\ref{u2}) is
to use the lowest approximation which neglects imaginary parts of
coefficients in the equations, but keeps the group-velocity terms. In this
approximation, the order parameters are real, $u_{1,2}(x)\equiv r_{1,2}(x)$,
obeying the time-independent version of Eqs. (\ref{u1}) and (\ref{u2}):
\begin{eqnarray}
+s\frac{dr_{1}}{dx} &=&\frac{d^{2}r_{1}}{dx^{2}}+r_{1}\left(
1-r_{1}^{2}-Gr_{2}^{2}\right) ,  \label{r11} \\
-s\frac{dr_{2}}{dx} &=&\frac{d^{2}r_{2}}{dx^{2}}+r_{2}\left(
1-r_{2}^{2}-Gr_{1}^{2}\right) .  \label{r22}
\end{eqnarray}%
Note that, unlike similar equations (\ref{r1}) and (\ref{r2}), Eqs. (\ref%
{r11}) and (\ref{r22}) cannot be derived from a formal Hamiltonian, cf. Eq. (%
\ref{h}).

\subsubsection{A (new) exact analytical solution}

To illustrate the structure of the DW state in this approximation, it is
relevant to produce a particular \emph{exact solution} of the system of
equations (\ref{r11}) and (\ref{r22}), cf. the above-mentioned solution (\ref%
{exact}):%
\begin{equation}
\left\{
\begin{array}{c}
r_{1}(x) \\
r_{2}(x)%
\end{array}%
\right\} =\frac{1}{2}\left\{
\begin{array}{c}
1-\tanh \left( \left( \sqrt{8+s^{2}}+s\right) (x/4\right)  \\
1+\tanh \left( \left( \sqrt{8+s^{2}}+s\right) (x/4\right)
\end{array}%
\right\} .  \label{exact3}
\end{equation}%
This solution (which was not reported in earlier works) exists if the
following relation holds between the cross-interaction coefficient $G$ and
group velocity $v$:%
\begin{equation}
G-3=s\left( \sqrt{8+s^{2}}+s\right) ,  \label{G-3}
\end{equation}%
or, inversely,%
\begin{equation}
s=\frac{G-3}{\sqrt{2\left( G+1\right) }}.  \label{v}
\end{equation}%
cf. Eq. (\ref{G=3}). It follows from the form of solution (\ref{exact3}) and
Eqs. (\ref{G-3}) and (\ref{v}) that
\begin{equation}
\mathrm{sgn}(s)=\mathrm{sgn}\left( G-3\right) ,  \label{sgn}
\end{equation}%
i.e., the exact solution represents a \textit{sink} of traveling waves ($s>0$%
) for $G>3$, and a \textit{source} ($s<0$) for $G<3$. Note that the solution
of the latter type exists even in the case of $G<1$, when the two components
are miscible, cf. Eq. (\ref{G>1}). In this case, the mixing is prevented by
the opposite group velocities, which pull the components apart.

\subsubsection{The sink or source coupled to a bright soliton in an
additional component}

It is possible to consider a system including the counterpropagating
traveling waves coupled to an additional standing one. This is a natural
counterpart of the three-component system based on Eqs. (\ref{u1g}), (\ref%
{u2g}), and (\ref{vv}). The traveling waves which can trap the additional
standing one, $v(x)$, are described by the following generalization of Eqs. (%
\ref{r11}) and (\ref{r22}):
\begin{eqnarray}
+s\frac{du_{1}}{dx} &=&\frac{d^{2}u_{1}}{dx^{2}}+u_{1}\left(
1-u_{1}^{2}-Gu_{2}^{2}-gv^{2}\right) ,  \label{ext1} \\
-s\frac{du_{2}}{dx} &=&\frac{d^{2}u_{2}}{dx^{2}}+u_{2}\left(
1-u_{2}^{2}-Gu_{1}^{2}-gv^{2}\right) ,  \label{ext2}
\end{eqnarray}%
while the equation for the standing component is
\begin{equation}
\frac{d^{2}v}{dx^{2}}+\left( 1-v^{2}-g\left( u_{1}^{2}+u_{2}^{2}\right)
\right) v=0,  \label{ext3}
\end{equation}%
cf. Eq. (\ref{vv}). An exact solution of Eqs. (\ref{ext1})-(\ref{ext3}) can
be found for free parameters $g$ and $s$:%
\begin{eqnarray}
u_{1,2}(x) &=&\frac{1}{2}\left( 1\mp \tanh \left( \sqrt{g-1}x\right) \right)
,  \label{ext4} \\
v(x) &=&\sqrt{2-\frac{3}{2}g}\mathrm{sech}\left( \sqrt{g-1}x\right) ,
\label{ext5}
\end{eqnarray}%
\begin{eqnarray}
G-3 &=&2g\left( 3g-4\right) +4s\sqrt{g-1},  \label{ext6} \\
D &=&\frac{s}{2\sqrt{g-1}}+\frac{1}{2}\left( 3g-1\right) ,  \label{ext7}
\end{eqnarray}%
cf. Eqs. (\ref{exact-v})-(\ref{DD}). As it is seen from Eq. (\ref{ext6}),
the interaction with the soliton-shaped standing wave shifts the boundary
between the sink and source of the traveling waves off the above-mentioned
point, $G=3$.

\section{Two- and three-dimensional quasiperiodic patterns}

Quasicrystals, as stable 3D\ ordered states of metallic alloys whose atomic
lattice is spatially quasiperiodic (QP), were discovered by D. Shechtman
\textit{et al}. \cite{Shechtman}. For this discovery, Shechtman was awarded
with the Nobel Prize in chemistry (2011). Then, a 2D quasicrystalline
structure was also experimentally demonstrated in alloys \cite{2D quasi}.
The work on this topic remains very active in diverse branches of
condensed-matter physics \cite{QC1,QC2,QC3}, as well as in other physical
systems which offer a natural realization of QP patterns, such as photonics
\cite{QC-opt0,QC-opt0,QC-opt2,QC-opt3} and phononics \cite{phononics}.

The objective of this section is to summarize results for stable 2D and 3D
patterns with the quasicrystalline structure that were predicted, as stable
non-equilibrium dynamical structures (rather than equilibrium states of
matter), in nonlinear dissipative media.

\subsection{2D octagonal (eight-mode) and decagonal (ten-mode) quasicrystals}

Following Ref. \cite{Trib-quasi}, generic 2D patterns of a real order
parameter, such as one representing the convection flow, are defined by
means of complex amplitudes $r_{l}$ of PWs which build them, cf. Eq. (\ref{U}%
):%
\begin{equation}
U\left( x,y,t\right) =\sum_{l=1}^{2N}u_{l}(t)\exp \left( i\mathbf{n}_{l}%
\mathbf{\cdot R}\right) ,  \label{sum}
\end{equation}%
where the set of $2N$ vectors $\mathbf{n}_{l}$ is a star with angles $\pi /N$
between adjacent ones. Note that the vectors satisfy the relation
\begin{equation}
\mathbf{n}_{l+N}=-\mathbf{n}_{l},~\mathrm{for}~l=1,2,...,N.  \label{nn}
\end{equation}%
Amplitudes $u_{l}(t)$ are, generally speaking, complex variables,%
\begin{equation}
u_{l}(t)=A_{l}(t)\exp \left( i\varphi _{l}(t)\right)  \label{Madelung2}
\end{equation}%
(cf. Eq. (\ref{Madelung})), subject to constraint $u_{l+N}=u_{l}^{\ast }$,
which, along with Eq. (\ref{nn}), secures that the order-parameter
distribution (\ref{sum}) is real.

For the lowest-order quasi-crystalline patterns, such as those corresponding
to $N=4$ (octagonal) and $N=5$ (decagonal) ones, the phase evolution is
trivial, making it possible to disregard $\varphi _{l}$ in Eq. (\ref%
{Madelung2}). The resulting system of evolution equations for the real
amplitudes, including the usual linear gain, $\gamma _{0}>0$, and cubic loss
(cf. Eq. (\ref{realGL})), is%
\begin{equation}
\frac{dA_{l}}{dt}=\left( \gamma _{0}-\sum_{m=1}^{N}T_{l-m}A_{m}^{2}\right)
A_{l}\equiv -\frac{\partial L}{\partial A_{l}},  \label{gamma0}
\end{equation}%
where $T_{l-m}>0$ are coefficients of the cubic lossy nonlinearity, subject
to normalization $T_{0}=1$, and the Lyapunov function is%
\begin{equation}
L=-\frac{\gamma _{0}}{2}\sum_{l=1}^{N}A_{l}^{2}+\frac{1}{4}%
\sum_{l,m=1}^{N}T_{l-m}A_{l}^{2}A_{m}^{2},  \label{L}
\end{equation}%
cf. Eqs. (\ref{grad}) and (\ref{Lyapunov}). Detailed analysis of the
results, produced with the help of Eq. (\ref{gamma0}), was presented in Ref.
\cite{Trib-quasi}), see also some preliminary results in Refs. \cite{MT} and
\cite{MNT-quasi}.

\begin{figure}[tbp]
\begin{center}
\subfigure[]{\includegraphics[width=0.45\textwidth]{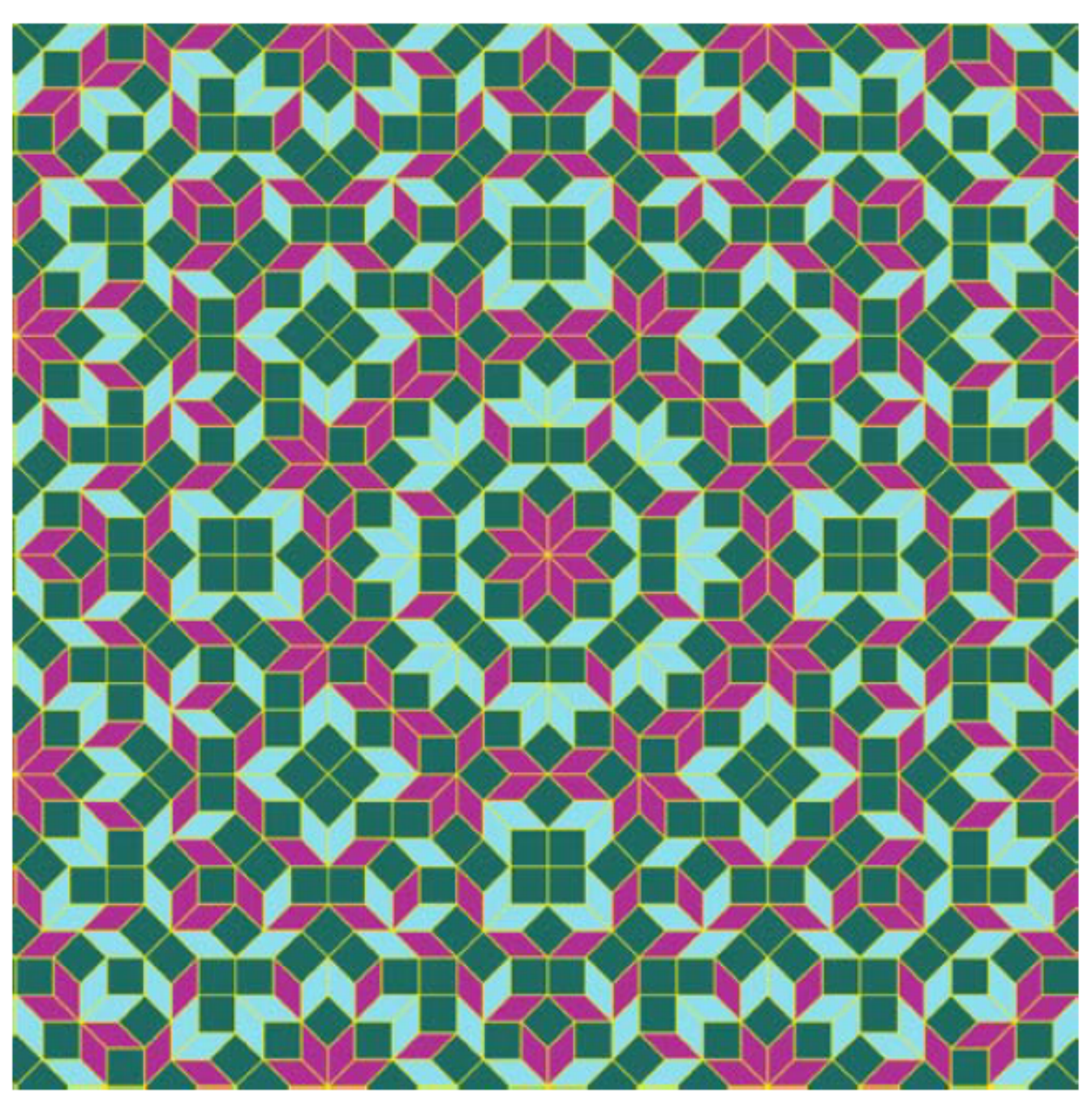}}%
\subfigure[]{\includegraphics[width=0.38\textwidth]{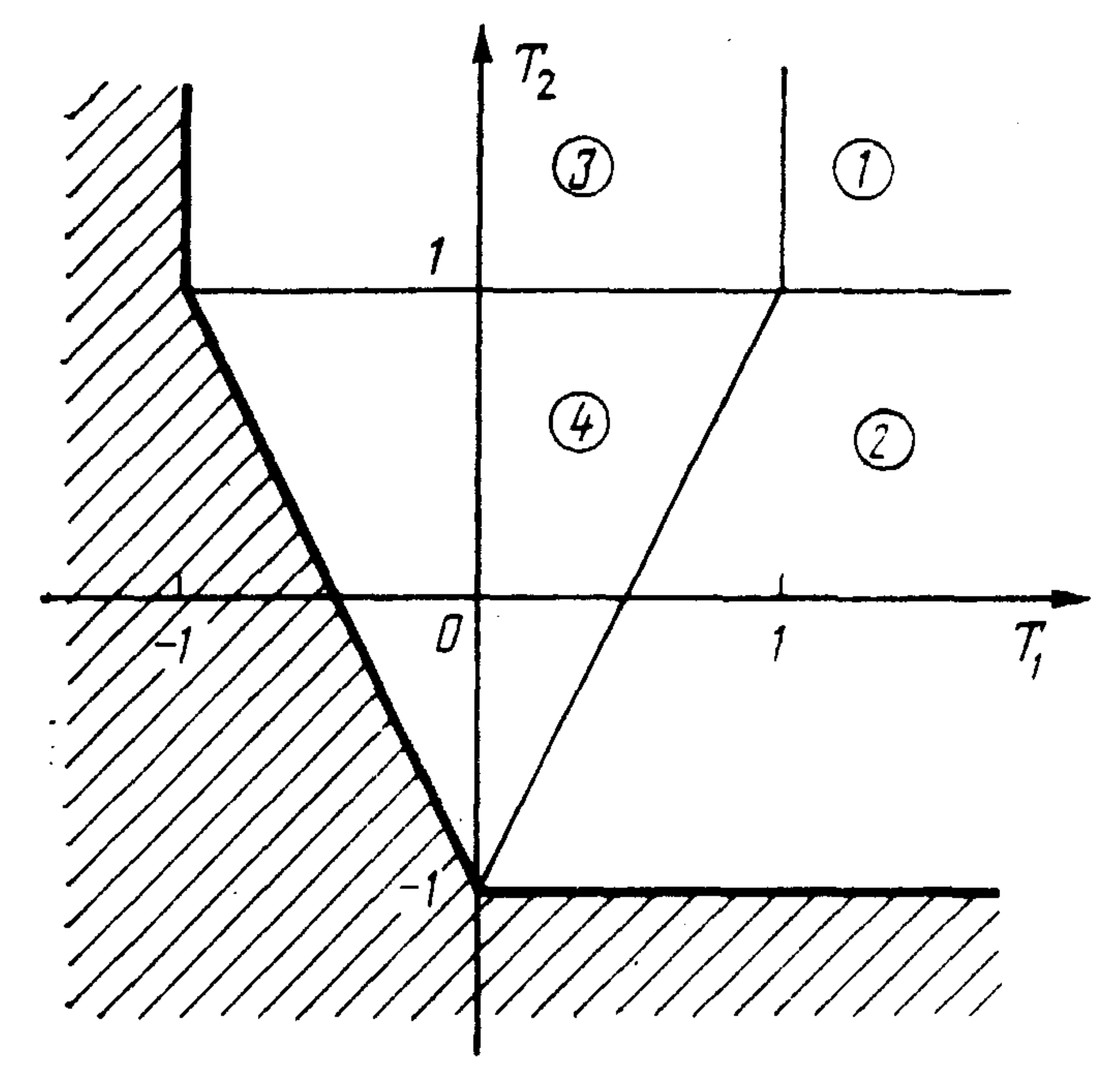}}
\end{center}
\caption{(a) The shape of the octagonal quasiperiodic pattern, reprinted
from Ref. \protect\cite{pictures}. (b) The stability chart for patterns
composed of four amplitudes, $A_{1,2,3,4}$, in the plane of nonlinearity
coefficients, $T_{1}$ and $T_{2}$, of Eq. (\protect\ref{gamma0}), reprinted
from Ref. \protect\cite{Trib-quasi}. Stability areas of the rolls (\protect
\ref{rolls}), squares (\protect\ref{squares}), rectangles (\protect\ref{rect}%
), and octagonal quasicrystal (\protect\ref{octag}) are denoted by encircled
numbers 1, 2, 3, and 4, respectively.}
\label{fig3}
\end{figure}

Spatially quasiperiodic patterns of the octagonal ($N=4$) and decagonal ($%
N=5 $) types are displayed, respectively, in Figs. \ref{fig3}(a) and \ref%
{fig4}(a). It is seen that they are built as compositions of rhombuses of
different shapes, and the presence of the overall octagonal or pentagonal
structure is evident.

Solutions of Eq. (\ref{gamma0}) for $N=4$ depend on two independent
nonlinearity coefficients, $T_{1}=T_{3}$ and $T_{2}$. In this case, there
are four distinct stationary solutions which have their stability areas:
rolls, with%
\begin{equation}
A_{1}=\sqrt{\gamma _{0}},A_{2,3,4}=0;  \label{rolls}
\end{equation}%
a square lattice, with%
\begin{equation}
A_{1,3}=\sqrt{\gamma _{0}/\left( 1+T_{2}\right) },A_{2,4}=0;  \label{squares}
\end{equation}%
an anisotropic rectangular lattice, with the aspect ratio $\tan (\pi
/8)=\allowbreak \sqrt{2}-1\approx \allowbreak 0.414$ and amplitudes%
\begin{equation}
A_{1,2}=\sqrt{\gamma _{0}/\left( 1+T_{1}\right) },A_{3,4}=0;  \label{rect}
\end{equation}%
and the octagonal quasicrystal, with%
\begin{equation}
A_{1,2,3,4}=\sqrt{\gamma _{0}/\left( 1+2T_{1}+T_{2}\right) }.  \label{octag}
\end{equation}

In addition to that, Eq. (\ref{gamma0}) also has a stationary semi-periodic
solution, which is quasiperiodic in direction $\mathbf{n}_{2}$ and periodic
along $\mathbf{n}_{4}$, with $A_{1}=A_{2}\neq 0,A_{3}\neq 0,$ and $A_{4}=0$.
However, the latter solution is completely unstable.

The full stability chart for stationary solutions (\ref{rolls})-(\ref{octag}%
) can be readily found in an analytical form. It is displayed in Fig. \ref%
{fig3}(b).

For $N=5$, solutions of Eq. (\ref{gamma0}) also depend on two independent
nonlinearity coefficients, $T_{1}=T_{4}$ and $T_{2}=T_{3}$. These equations
produce six different species of stable stationary patterns: rolls, with%
\begin{equation}
A_{1}=\sqrt{\gamma _{0}},A_{2,3,4,5}=0;  \label{rolls2}
\end{equation}%
two different species of rectangular lattices:%
\begin{eqnarray}
A_{1,2} &=&\sqrt{\gamma _{0}/\left( 1+T_{1}\right) },A_{3,4,5}=0,
\label{rect2} \\
A_{1,3} &=&\sqrt{\gamma _{0}/\left( 1+T_{2}\right) },A_{2,4,5}=0;
\label{rect3}
\end{eqnarray}%
the decagonal quasicrystal:%
\begin{equation}
A_{1,2,3,4,5}=\sqrt{\gamma _{0}/\left( 1+2T_{1}+2T_{2}\right) };
\label{decag}
\end{equation}%
and two species of semi-periodic patterns, that are quasiperiodic in one
direction and periodic in the other:%
\begin{eqnarray}
A_{1,3} &=&\sqrt{\frac{\gamma _{0}\left( 1-T_{1}\right) }{1+T_{2}-2T_{1}^{2}}%
},A_{2}=\sqrt{\frac{\gamma _{0}\left( 1+T_{2}-2T_{1}\right) }{%
1+T_{2}-2T_{1}^{2}}},A_{4,5}=0,  \label{mixed1} \\
A_{1,5} &=&\sqrt{\frac{\gamma _{0}\left( 1-T_{2}\right) }{1+T_{1}-2T_{2}^{2}}%
},A_{3}=\sqrt{\frac{\gamma _{0}\left( 1+T_{1}-2T_{2}\right) }{%
1+T_{1}-2T_{2}^{2}}},A_{2,4}=0.  \label{mixed2}
\end{eqnarray}%
In addition to that, there is another semi-periodic solution, with $%
A_{1}=A_{2}\neq 0$, $A_{3}=A_{4}\neq 0$, and $A_{5}=0$, but it is completely
unstable.

\begin{figure}[tbp]
\begin{center}
\subfigure[]{\includegraphics[width=0.45\textwidth]{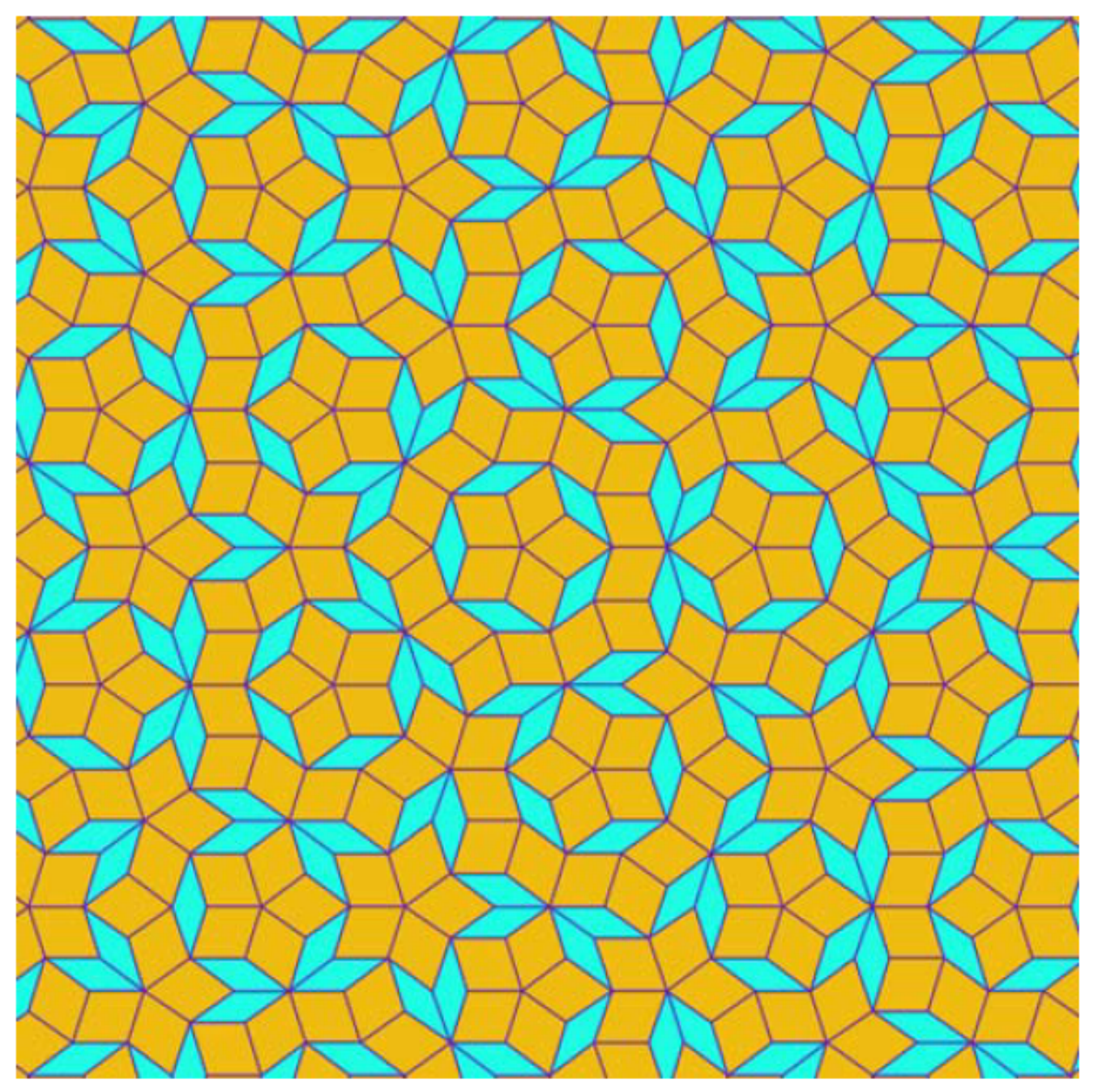}}%
\subfigure[]{\includegraphics[width=0.38\textwidth]{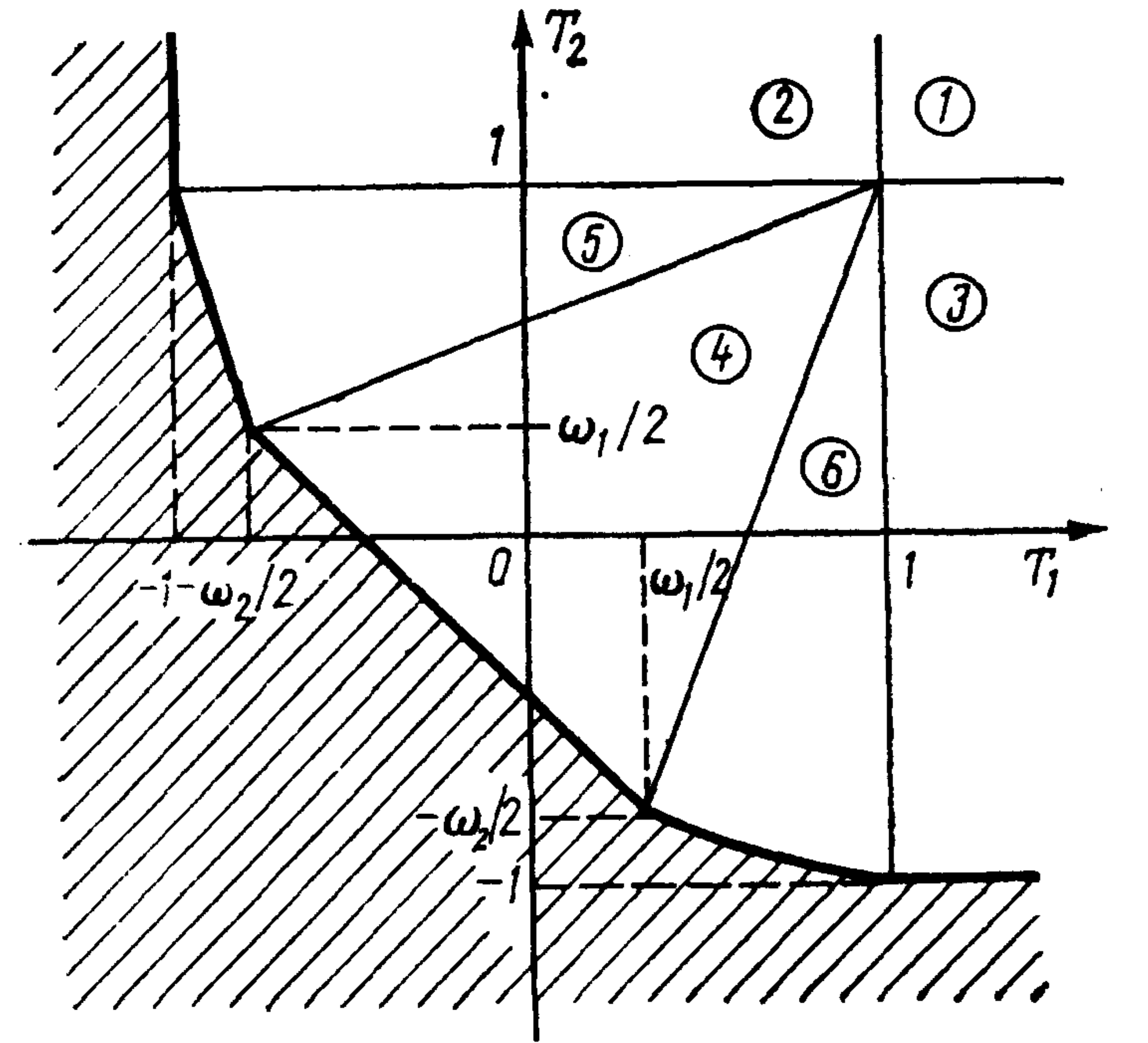}}
\end{center}
\caption{(a) The shape of the decagonal quasiperiodic pattern, reprinted
from Ref. \protect\cite{pictures}. (b) The stability chart for patterns
composed of five amplitudes, $A_{1,2,3,4,5}$, in the plane of nonlinearity
coefficients, $T_{1}$ and $T_{2}$, of the respective system of equations (%
\protect\ref{gamma0}), reprinted from Ref. \protect\cite{Trib-quasi}.
Stability areas of the rolls (\protect\ref{rolls2}), rectangles (\protect\ref%
{rect2}) and (\protect\ref{rect3}), decagonal quasicrystal (\protect\ref%
{decag}), and semi-periodic states (\protect\ref{mixed1}) and (\protect\ref%
{mixed2}), are denoted by encircled numbers 1, 2, 3, 4, 5, and 6,
respectively. Constants $\protect\omega _{1}$ and $\protect\omega _{2}$,
marked in panel (b), are defined by Eq. (\protect\ref{omega}).}
\label{fig4}
\end{figure}

The full stability chart for this set of solutions was found too in an
analytical form, as shown in Fig. \ref{fig4}(b). In this figure, the
constants are%
\begin{equation}
\omega _{1}=\frac{\sqrt{5}-1}{2}\approx 0.618,\omega _{2}=\omega _{1}+1.
\label{omega}
\end{equation}%
Note that, unlike the situation for the octagonal setting ($N=4$), displayed
in Fig. \ref{fig3}(b), the stability areas for the decagonal ($N=5$)
quasicrystal (\ref{decag}) and periodic patterns (\ref{rect2}), (\ref{rect3}%
), are not adjacent to each other in Fig. \ref{fig4}(b), being separated by
regions of stable semi-periodic states (\ref{mixed1}) and (\ref{mixed2})
(recall that all semi-periodic states are unstable in the case of $N=4$).

It is relevant to mention that, if higher-order nonlinear terms are added to
the system of equations (\ref{gamma0}), the sharp boundaries between
stability areas of different patterns in Figs. \ref{fig3}(b) and \ref{fig4}%
(b) may be modified. In particular, there may appear narrow strips of
bistability (which is impossible in the framework of Eq. (\ref{gamma0})), as
well as strips populated by more complex patterns, instead of the sharp
lines \cite{Trib-quasi}.

\subsection{Dodecagonal quasicrystals ($N=6$)}

In the case of the twelve-mode patterns, corresponding to $N=6$ in Eq. (\ref%
{gamma0}), quadratic nonlinearity, with coefficient $\nu >0$, should be
included too, as the corresponding set of six wave vectors contains two
\textit{resonant triads}, that may be naturally coupled by the quadratic
terms (cf. Eq. (\ref{triple})):%
\begin{equation}
\mathbf{n}_{1}+\mathbf{n}_{5}+\mathbf{n}_{9}=\mathbf{n}_{2}+\mathbf{n}_{6}+%
\mathbf{n}_{10}=0  \label{triads}
\end{equation}%
(recall that only six wave vectors are actually different, according to Eq. (%
\ref{nn})). In this case, the dynamics of phases of the complex amplitudes (%
\ref{Madelung}) cannot be disregarded, giving rise to \textit{phason modes}
\cite{phason1,phason2,Lifshitz}. Accordingly, Eq. (\ref{gamma0}) is replaced
by a coupled system of evolution equations for the real amplitudes and
phases \cite{Trib-quasi}:%
\begin{gather}
\frac{dA_{l}}{dt}=\left( \gamma _{0}-\sum_{m=1}^{N}T_{l-m}A_{m}^{2}\right)
A_{l}+\nu A_{n+4}A_{n+8}\cos \Phi _{n},  \label{dA/dt} \\
A_{l}\frac{d\varphi _{l}}{dt}=-\nu A_{l+4}A_{l+8}\sin \Phi _{l},
\label{dphi/dt} \\
\Phi _{l}\equiv \varphi _{l}+\varphi _{l+4}+\varphi _{l+8},  \label{Phi}
\end{gather}%
with $\varphi _{l}\equiv \varphi _{l-12}$ for $l>12$.

These equations give rise to the following stationary solutions for the
dodecagonal quasicrystals, with equal values of real amplitudes $A_{l}$, and
the same values of $\Phi _{l}$ for both resonantly coupled triads (\ref%
{triads}):%
\begin{gather}
\cos \Phi _{l}=\pm 1,  \label{cos} \\
A_{l}=\pm \left( 2Q_{0}\right) ^{-1}\left( \nu \pm \sqrt{\nu ^{2}+4\gamma
_{0}Q_{0}}\right) ,  \label{Al} \\
Q_{0}\equiv 1+2T_{1}+2T_{2}+T_{3}.  \label{Q0}
\end{gather}%
The analysis of the stability of these solutions in the framework of Eqs. (%
\ref{dA/dt})-(\ref{Phi}) demonstrates that they may be stable only under
conditions $Q_{0}>0$ (see Eq. (\ref{Q0}) and
\begin{equation}
Q_{3}\equiv 1+T_{3}-T_{1}-T_{2}>0.  \label{Q3}
\end{equation}%
If these conditions hold, the amplitude of stable quasicrystals exceeds a
minimum value,%
\begin{equation}
A_{l}\geq A_{\min }\equiv (1/2)\max \left\{ \nu /Q_{0},\nu /Q_{3}\right\} .
\label{Amin}
\end{equation}%
Further, there is no stability constraint from above for the amplitude,
provided that the following combinations of the nonlinearity coefficients
are positive:%
\begin{equation}
Q_{1,2}\equiv 1-T_{2}\pm \left( T_{1}-T_{3}\right) \geq 0.  \label{Q12}
\end{equation}%
Otherwise, the stability imposes the following limit on the amplitude:%
\begin{equation}
A\leq A_{\max }\equiv \min \left\{ -\nu /Q_{1},-\nu /Q_{2}\right\}
\label{Amax}
\end{equation}%
(if only one combination $Q_{1}$ or $Q_{2}$ is negative, only this one
determines the upper limit for the stability, as per Eq. (\ref{Amax})).

\begin{figure}[tbp]
\begin{center}
\subfigure[]{\includegraphics[width=0.45\textwidth]{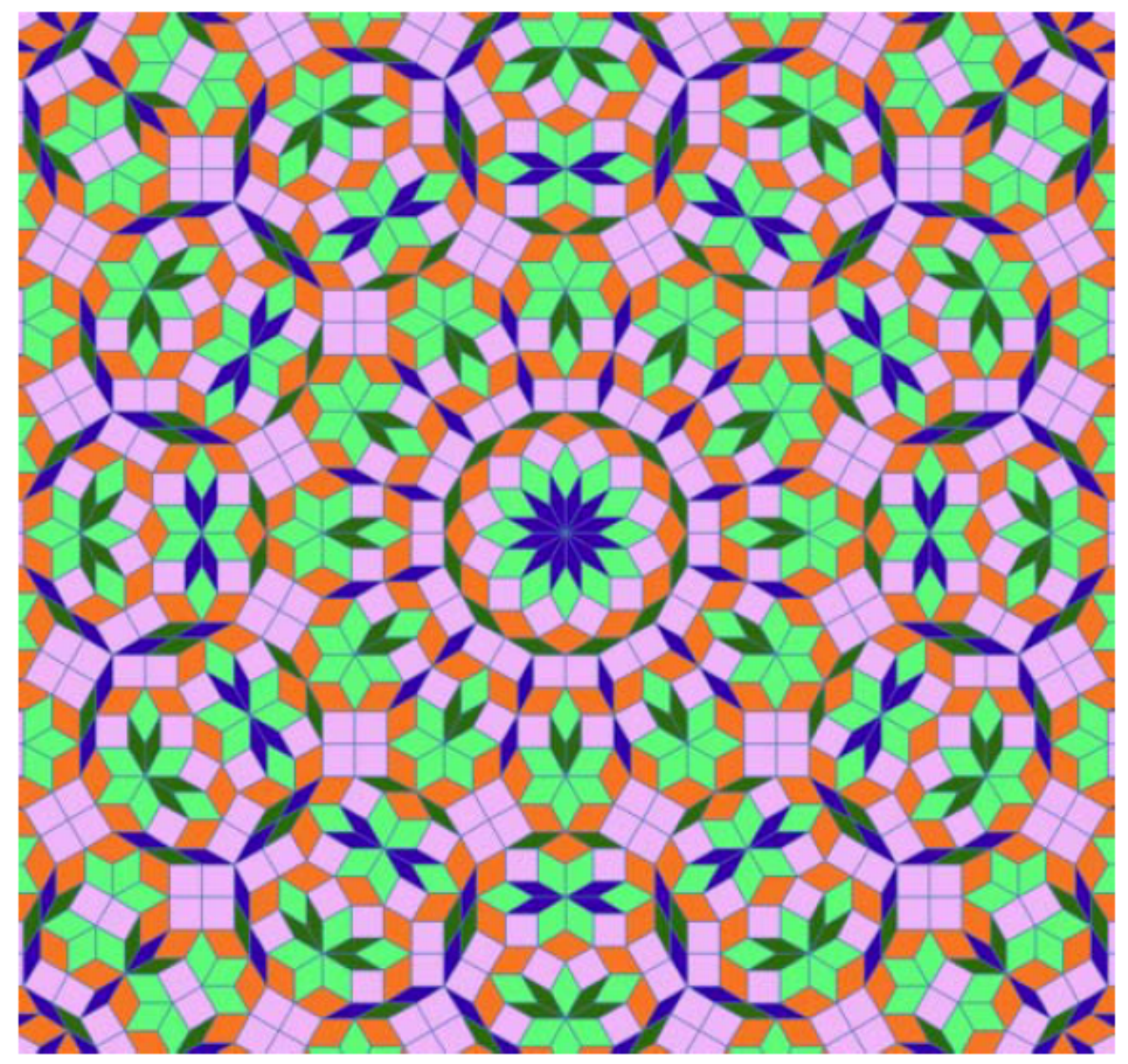}}%
\subfigure[]{\includegraphics[width=0.38\textwidth]{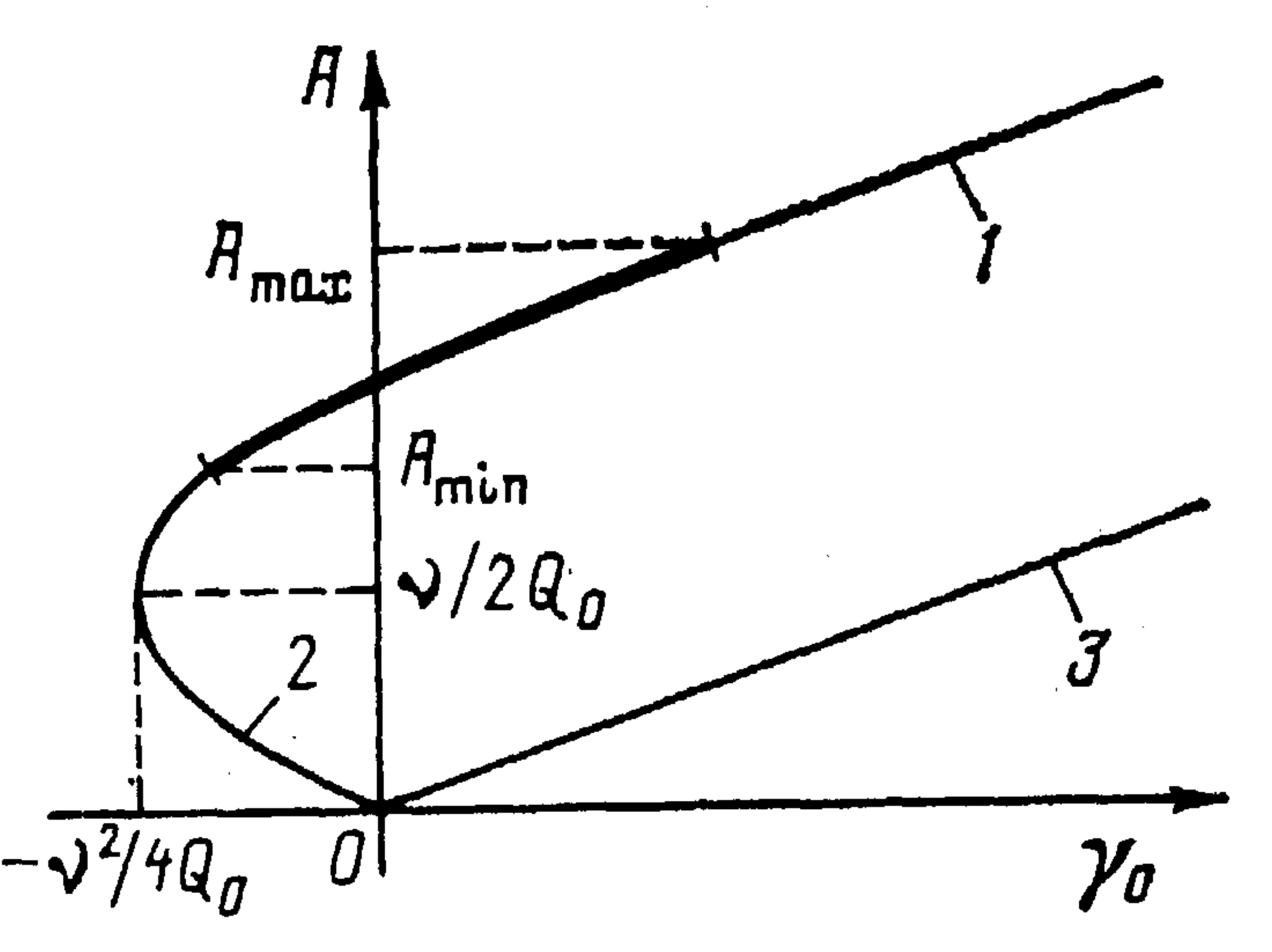}}
\end{center}
\caption{(a) The shape of the dodecagonal quasiperiodic pattern, reprinted
from Ref. \protect\cite{pictures}. (b) The pattern's amplitude, given by Eq.
(\protect\ref{Al}), vs. the strength of the linear gain ($\protect\gamma %
_{0}>0$) or loss ($\protect\gamma _{0}<0$), reprinted from Ref. \protect\cite%
{Trib-quasi}. Stable and unstable solutions are represented, respectively,
by bold and thin lines. Branches 1, 2 and 3 pertain, severally, to the
solutions with $\cos \Phi _{l}=+1$ and $-1$ in Eq. (\protect\ref{cos}).
Parameters $Q_{0}$, $A_{\min }$, and $A_{\max }$ are defined as per Eqs. (%
\protect\ref{Q0}), (\protect\ref{Amin}), and (\protect\ref{Amax}),
respectively.}
\label{fig5}
\end{figure}

The existence and stability results for the amplitude of the dodecagonal
quasicrystals is summarized in Fig. \ref{fig5}(b). Note that, in the
presence of the resonant interaction mediated by the quadratic term in Eq. (%
\ref{dA/dt}), the solution appears as a \textit{subcritical} \cite%
{subcritical} one, with a finite value of the amplitude, at $\gamma _{0}<0$,
i.e., when this coefficient represents linear \textit{loss}, rather than
gain.

\subsection{A quasicrystalline layer between orthogonally oriented
square-lattice patterns}

\begin{figure}[tbp]
\begin{center}
\subfigure[]{\includegraphics[width=0.46\textwidth]{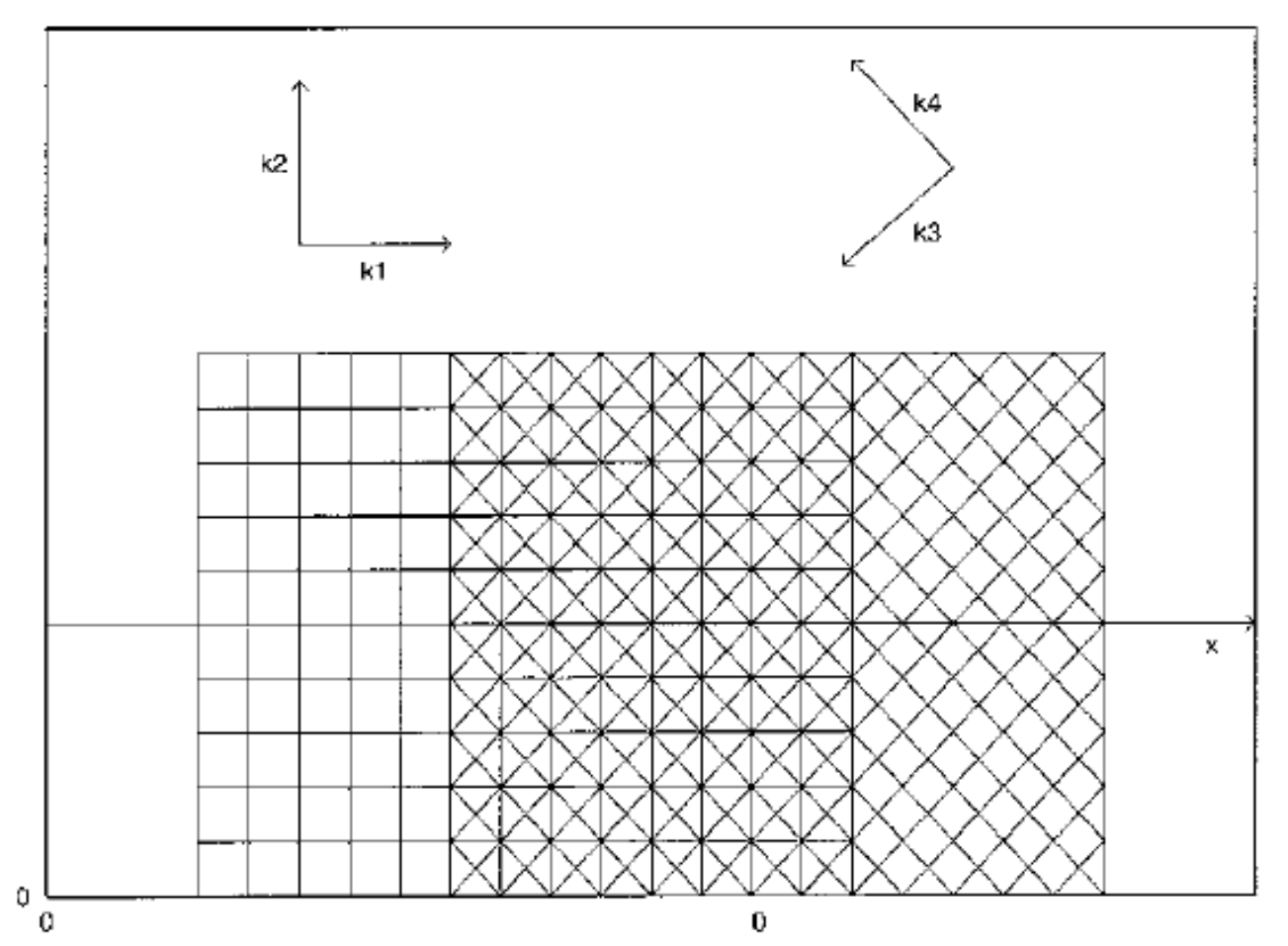}}%
\subfigure[]{\includegraphics[width=0.46\textwidth]{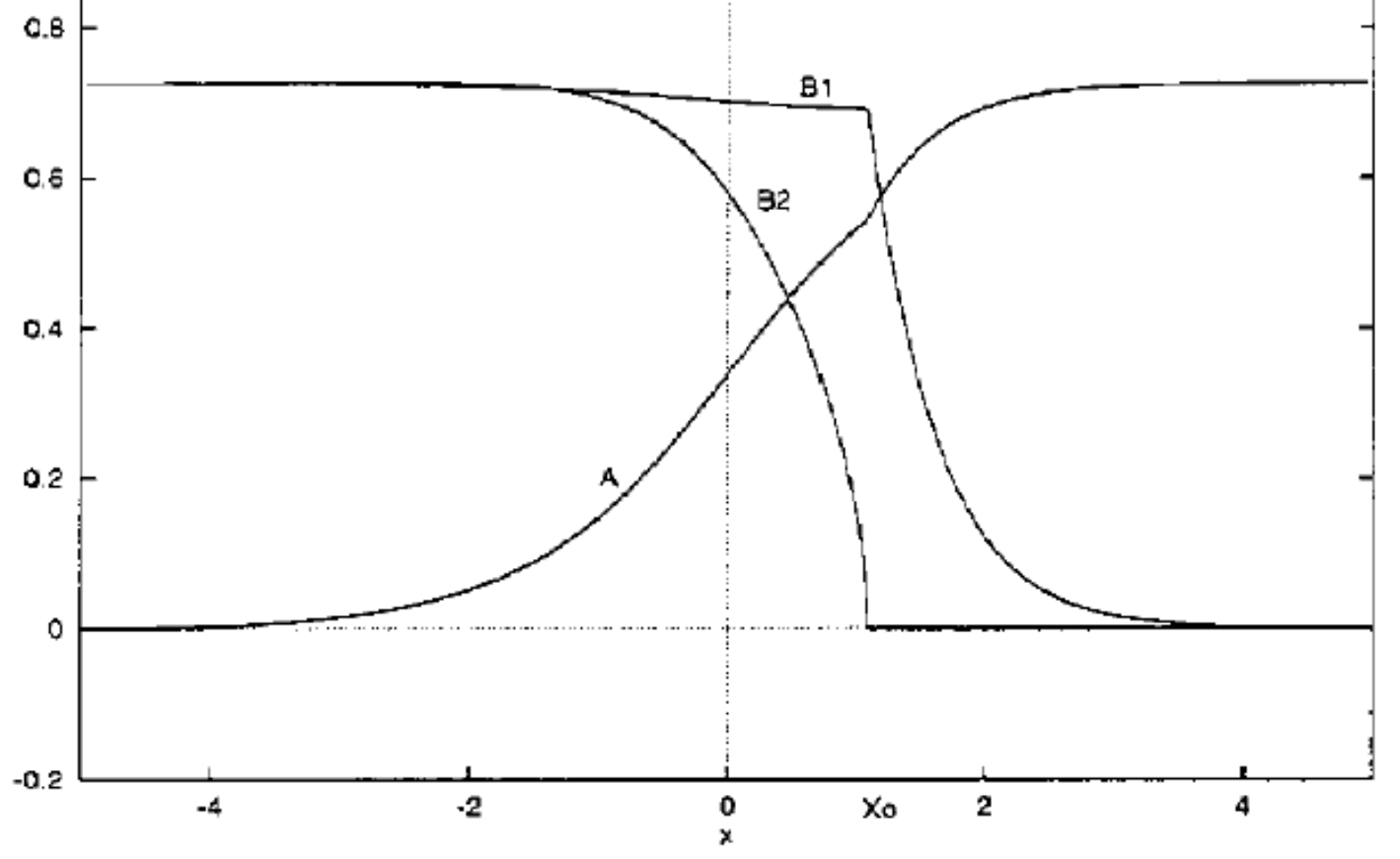}}
\end{center}
\caption{(a) The scheme for building a broad stripe of the octagonal
quasicrystalline state as a transient layer between semi-infinite domains
filled by square-lattice patterns, mutually rotated by $45^{\mathrm{o}}$.
(b) An example of the corresponding solution for amplitudes $A(x)$ and $%
B_{1,2}(x)$. Parts of the solution corresponding to Eq. (\protect\ref{TF2})
or to $B_{2}=0$ are connected at the stitch point $x=x_{0}$. Reprinted from
Ref. \protect\cite{Rotstein}.}
\label{fig6}
\end{figure}

While, as shown in Fig. \ref{fig3}(b), square-lattice and octagonal
quasiperiodic patterns cannot coexist as stable ones in the system with $N=4$%
, it was demonstrated in Ref. \cite{Rotstein} that a sufficiently broad
stripe filled by an effectively stable quasiperiodic pattern may be realized
as a transient layer between stable square-lattice patterns mutually
oriented under the angle of $45^{\mathrm{o}}$, as schematically shown in
Fig. \ref{fig6}(a). For this configuration (which naturally combines the two
main topics of the present article, \textit{viz}., the DWs and QP patterns),
one may naturally adopt equal amplitudes corresponding to wave vectors $%
\mathbf{k}_{3}$ and $\mathbf{k}_{4}$,
\begin{equation}
A_{3}(x)=A_{4}(x)\equiv A(x),  \label{34}
\end{equation}%
while amplitudes $B_{1}$ and $B_{2}$ related to $\mathbf{k}_{1}$ and $%
\mathbf{k}_{2}$ are different, the effective diffusion coefficient for the
latter one being zero, as per Eq. (\ref{D}). The corresponding system of
stationary real GL equations, naturally extending Eqs. (\ref{lambda1}), (\ref%
{lambda2}), (\ref{D1}), (\ref{D=0}), and (\ref{gamma0}), takes the following
form:%
\begin{gather}
\frac{1}{2}\frac{d^{2}A}{dx^{2}}+A-A^{3}-\left(
T_{1}B_{1}^{2}+T_{1}B_{2}^{2}+T_{2}A^{2}\right) A=0,  \label{A3=A4} \\
\frac{d^{2}B_{1}}{dx^{2}}+B_{1}-B_{1}^{3}-\left(
2T_{1}A^{2}+T_{2}B_{2}^{2}\right) B_{1}=0,  \label{B1} \\
B_{2}-B_{2}^{3}-\left( 2T_{1}A^{2}+T_{2}B_{1}^{2}\right) B_{2}=0,  \label{B2}
\end{gather}%
where, like in Eq. (\ref{gamma0}), $T_{1}$ and $T_{2}$ are coefficients of
the cross-interaction between the PW modes with angles, respectively, $45^{%
\mathrm{o}}$ and $90^{\mathrm{o}}$ between their wave vectors. According to
Fig. \ref{fig3}(b), the stability conditions for the spatially uniform
square-lattice and octagonal quasicrystalline patterns are, respectively,%
\begin{eqnarray}
T_{2} &\leq &1,T_{1}\geq T_{2}+1/2,  \label{T2T1} \\
T_{2} &\leq &1,T_{1}\leq T_{2}+1/2.  \label{T1T2}
\end{eqnarray}%
Accordingly, to secure the stability of the background square lattices and a
possibility to have a broad transient layer between mutually rotated ones,
which is filled by the effectively stable octagonal pattern, it is relevant
to choose parameters belonging to the stability area (\ref{T2T1}), with
values close to the stability boundary, $T_{1}=T_{2}+1/2$. An appropriate
choice is%
\begin{equation}
T_{1,2}\equiv 1-\mu _{1,2},~0<\mu _{1,2}\ll 1,  \label{T12}
\end{equation}%
\begin{equation}
\text{with }m\equiv 2\mu _{1}/\mu _{2}<1.  \label{m}
\end{equation}

Similar to what is considered above in Eqs. (\ref{r2=0}) and (\ref{TF}), Eq.
(\ref{B2}) obviously splits in two options: $B_{2}=0$, or%
\begin{equation}
B_{2}^{2}+2T_{1}A^{2}+T_{2}B_{1}^{2}=1.  \label{TF2}
\end{equation}%
In either case, Eqs. (\ref{A3=A4}) and (\ref{B1}) simplify accordingly. The
solutions corresponding to $B_{2}=0$ or to Eq. (\ref{TF2}) must be
\textquotedblleft dovetailed" at a stitch point $x=x_{0}$, cf. Eq. (\ref{x0}%
). An example of the so obtained solutions for amplitudes $A(x)$ and $%
B_{1,2}(x)$ is displayed in Fig. \ref{fig6}(b).

\subsection{Three-dimensional quasicrystals}

A setting which makes it possible to predict a stable quasiperiodic pattern,
based on a set of four PW modes, in the 3D space was put forward in Ref.
\cite{Komarova}. It originates from the model of a lasing cavity, based on
the standard system of coupled Maxwell-Bloch equations. The evolutional
variable in this system is time, while the spatial structure is strongly
anisotropic, as the field (Maxwell's) equation in the system contains only
the first derivative, $\partial /\partial z$, with respect to the
longitudinal coordinate, $z$, and the usual paraxial-diffraction operator, $%
i\left( \partial _{x}^{2}+\partial _{y}^{2}\right) ,$acting on the
transverse coordinates, $\left( x,y\right) $. As a result, at the lasing
threshold components of 3D wave vectors carrying the PW modes,
\begin{equation}
\mathbf{K}=(\mathbf{k},k_{z}),~\mathbf{k}\equiv \left( k_{x},k_{y}\right) ,
\label{kq}
\end{equation}%
satisfy the following dispersion relation, which couples them to the wave's
frequency $\Omega $:%
\begin{equation}
\Omega =k^{2}+k_{z}.  \label{Omega}
\end{equation}

Eventually, above the lasing threshold the cubic nonlinearity of the
Maxwell-Bloch system may produce a \emph{resonant quartet} of 3D wave
vectors, coupled by condition%
\begin{equation}
\mathbf{K}_{1}+\mathbf{K}_{2}=\mathbf{K}_{3}+\mathbf{K}_{4}.  \label{1234}
\end{equation}%
For comparison, in the 2D space the same relation (\ref{1234}), taken close
to the threshold, i.e., for nearly equal length of the wave vectors, would
imply that the four vectors form a rhombus, and the cubic interaction
between the corresponding amplitudes, $u_{1,2,3,4}$, would be represented by
usual nonresonant nonlinear terms, essentially the same as in Eq. (\ref%
{gamma0}), with $A_{l}$ replaced by $u_{l}$ and $A_{m}^{2}A_{l}$ replaced by
the XPM terms, $\left\vert u_{m}\right\vert ^{2}u_{l}$. However, in the 3D
setting the resonance condition (\ref{1234}), combined with the dispersion
relation (\ref{Omega}), leads to a nontrivial possibility to add \textit{%
four-wave-mixing} (FWM) cubic terms to the XPM ones, see below.

\begin{figure}[tbp]
\begin{center}
\subfigure[]{\includegraphics[width=0.42\textwidth]{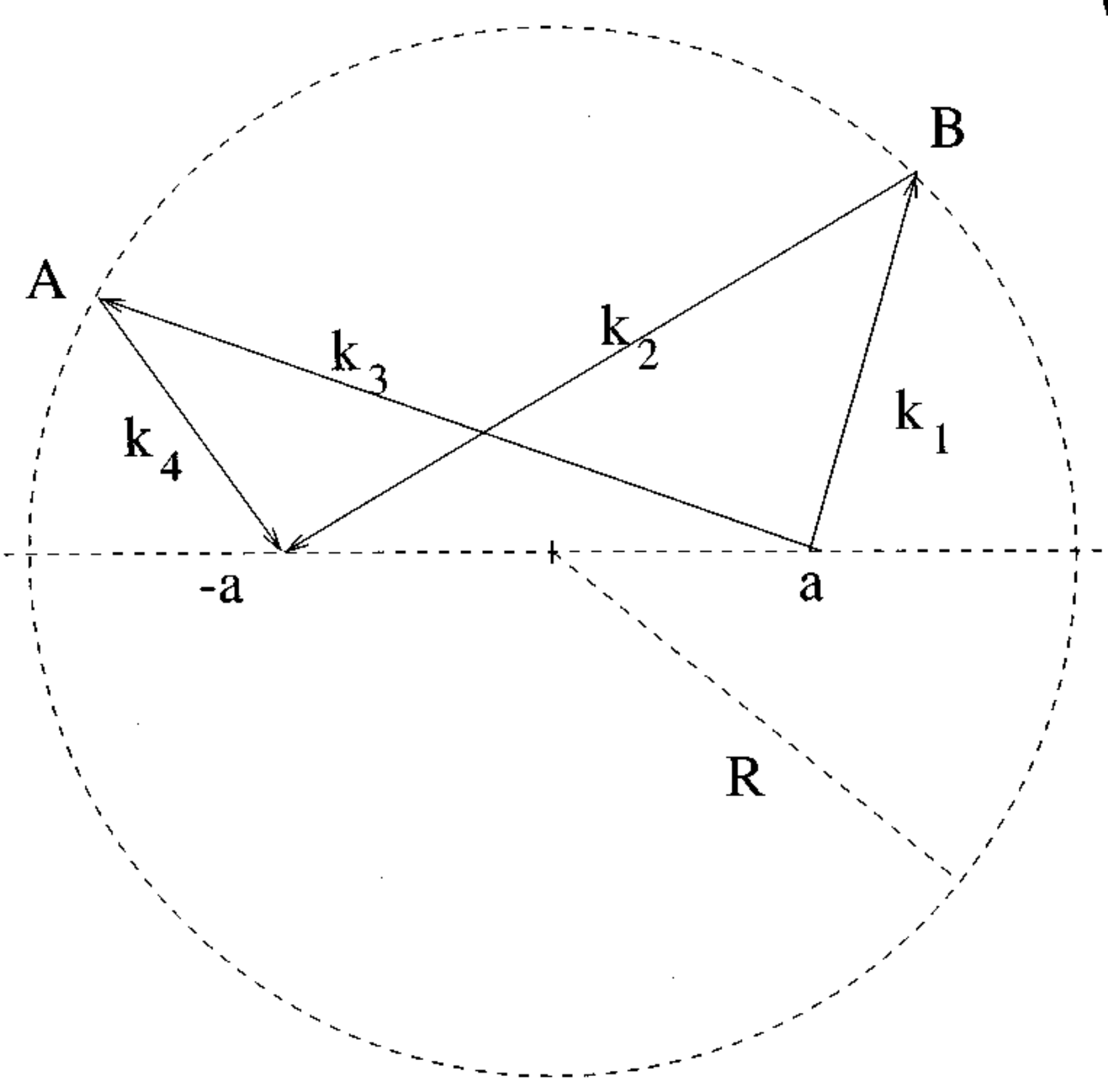}}%
\subfigure[]{\includegraphics[width=0.46\textwidth]{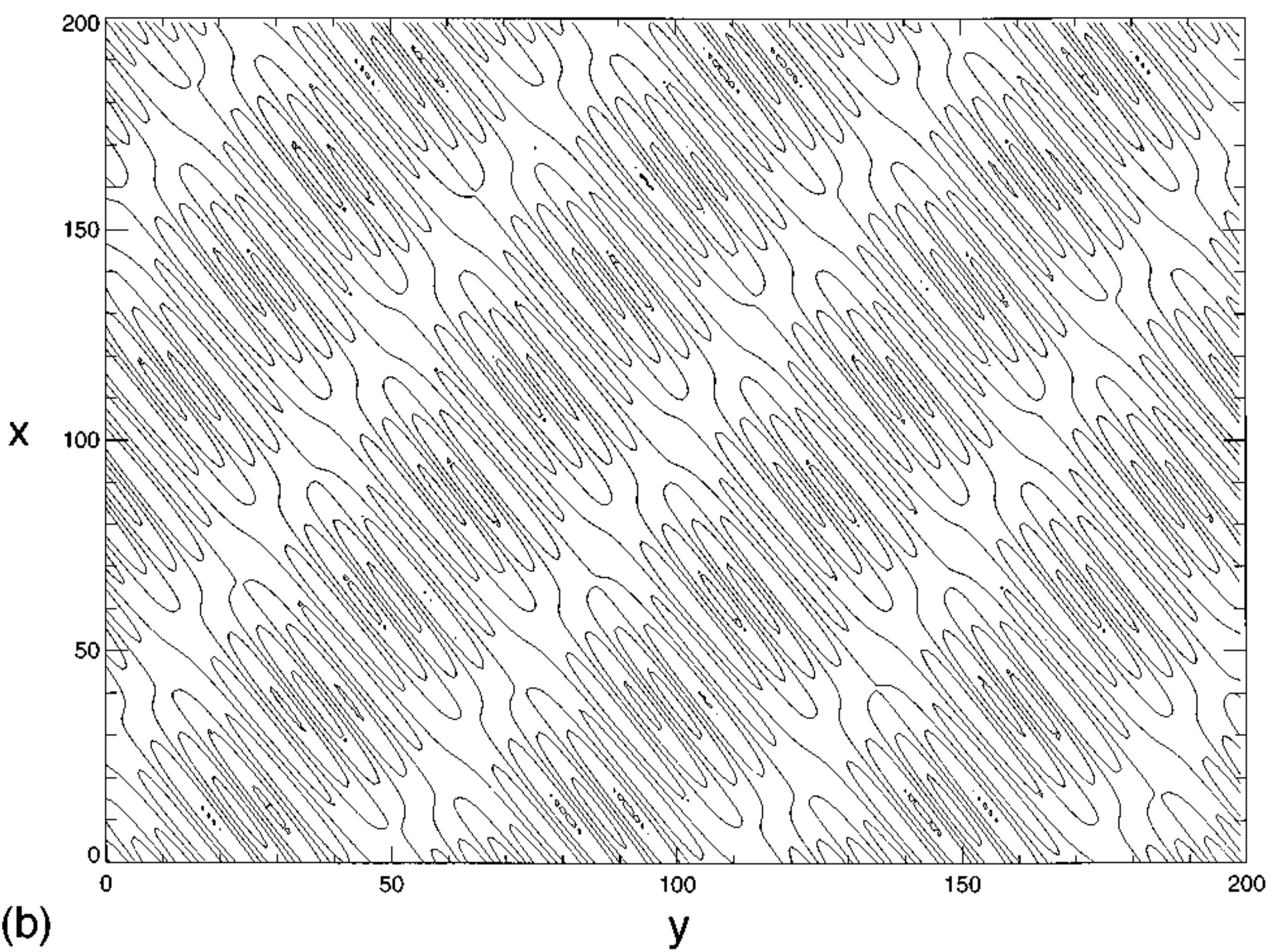}}
\end{center}
\caption{(a) A set of four two-dimensional vectors $\mathbf{k}_{1,2,3,4}$
which solves equations (\protect\ref{geometry}). Here, A and B are two
arbitrary points belonging to the circumference of arbitrary radius $R$, and
$a<R$ is an arbitrary value of coordinate $x$. (b) An example of the
three-dimensional quasiperiodic pattern, projected onto the $\left(
x,y\right) $ plane. Shown are contour plots of the corresponding
distribution of the order parameter, $\mathrm{Re}\left[ \sum_{l=1}^{4}u_{l}%
\cos \left( \mathbf{k}_{l}\cdot \mathbf{R}\right) \right] $, where $u_{l}$
are the complex amplitudes given by Eqs. (\protect\ref{3D}), (\protect\ref%
{pi}), and (\protect\ref{1/5}). In this case, the phases are $\protect%
\varphi _{1}=\protect\pi ,\protect\varphi _{2}=-\protect\pi /2,$ $\protect%
\varphi _{3}=\protect\pi /7$, while $\protect\varphi _{4}$ is determined by
Eq. (\protect\ref{pi}). The angle between vectors $\mathbf{k}_{1}-\mathbf{k}%
_{2}$ and $\mathbf{k}_{3}-\mathbf{k}_{4}$ is $\protect\pi /5$. The figure is
reprinted from Ref. \protect\cite{Komarova}.}
\label{fig7}
\end{figure}

Substituting expression (\ref{kq}) for the 3D wave vector in Eqs. (\ref{1234}%
) and (\ref{Omega}) leads to the following elementary exercise in planar
geometry: find two pairs of 2D vectors, $\left( \mathbf{k}_{1},\mathbf{k}%
_{2}\right) $ and $\left( \mathbf{k}_{3},\mathbf{k}_{4}\right) $, satisfying
conditions%
\begin{equation}
\mathbf{k}_{1}+\mathbf{k}_{2}=\mathbf{k}_{3}+\mathbf{k}%
_{4},~k_{1}^{2}+k_{2}^{2}=k_{3}^{2}+k_{4}^{2}.  \label{geometry}
\end{equation}%
An obvious solution of this exercise is plotted in Fig. \ref{fig7}(a).

Once a resonantly coupled quartet of four wave vectors is chosen, the
respective system of evolution equations for the corresponding complex
amplitudes is \cite{Komarova}%
\begin{eqnarray}
\frac{du_{1}}{dt} &=&\gamma _{0}u_{1}-\left( \left\vert u_{1}\right\vert
^{2}+2\sum_{l\neq 1}\left\vert u_{l}\right\vert ^{2}\right)
u_{1}-2u_{2}^{\ast }u_{3}u_{4},  \notag \\
\frac{du_{2}}{dt} &=&\gamma _{0}u_{2}-\left( \left\vert u_{2}\right\vert
^{2}+2\sum_{l\neq 2}\left\vert u_{l}\right\vert ^{2}\right)
u_{2}-2u_{1}^{\ast }u_{3}u_{4},  \notag \\
&&  \label{FWM} \\
\frac{du_{3}}{dt} &=&\gamma _{0}u_{3}-\left( \left\vert u_{3}\right\vert
^{2}+2\sum_{l\neq 3}\left\vert u_{l}\right\vert ^{2}\right)
u_{3}-2u_{1}u_{2}u_{4}^{\ast },  \notag \\
\frac{du_{4}}{dt} &=&\gamma _{0}u_{4}-\left( \left\vert u_{4}\right\vert
^{2}+2\sum_{l\neq 4}\left\vert u_{l}\right\vert ^{2}\right)
u_{4}-2u_{1}u_{2}u_{3}^{\ast }.  \notag
\end{eqnarray}%
In these equations, $\gamma _{0}>0$ is the linear gain, as above, and the
last terms represent the above-mentioned FWM effect. Particular values of
coefficients in front of nonlinear terms are standard ones which correspond
to the XPM and FWM interactions in nonlinear optics \cite{KM}, unlike
general values of coefficients $T_{l-m}$ in Eq. (\ref{gamma0}). Similar to
Eq. (\ref{gamma0}), the system of equations (\ref{FWM}) admits the
presentation in the form of $du_{l}/dt=-$ $\partial L/\partial u_{l}^{\ast }$%
, with the Lyapunov function%
\begin{equation}
L=-\gamma _{0}\sum_{l}\left\vert u_{l}\right\vert ^{2}+\frac{1}{2}%
\sum_{l}\left\vert u_{l}\right\vert ^{4}+2\sum_{l>m}\left\vert
u_{l}\right\vert ^{2}\left\vert u_{m}\right\vert ^{2}+4\mathrm{Re}\left(
u_{1}u_{2}u_{3}^{\ast }u_{4}^{\ast }\right) .  \label{LL}
\end{equation}

Further analysis performed in Ref. \cite{Komarova} had produced two stable
stationary solutions of Eqs. (\ref{FWM}). First, this is a simple
single-mode state (rolls), with%
\begin{equation}
\left\vert u_{1}\right\vert ^{2}=\gamma _{0},~u_{2,3,4}=0.
\label{single-mode}
\end{equation}%
Next, dodecagonal quasicrystals with equal absolute values of all the four
amplitudes are looked for as%
\begin{equation}
u_{l}=A\exp \left( i\varphi _{l}\right) ,  \label{3D}
\end{equation}%
where the phased are locked so that%
\begin{equation}
\varphi _{1}+\varphi _{2}-\varphi _{3}-\varphi _{4}=\pi ,  \label{pi}
\end{equation}%
and the squared absolute value of the amplitudes is%
\begin{equation}
\left\vert u_{l}\right\vert ^{2}=\gamma _{0}/5,  \label{1/5}
\end{equation}%
cf. the rolls solution (\ref{single-mode}). Note that values of the Lyapunov
function (\ref{LL}) for the rolls and 3D quasicrystal are%
\begin{equation}
L_{\mathrm{rolls}}=-\gamma _{0}^{2}/2,~L_{\mathrm{quasicryst}}=-2\gamma
_{0}^{2}/5,  \label{LLL}
\end{equation}%
hence the rolls represent the ground state of the system, while the
quasicrystal is a metastable state, as its value of $L$ is slightly higher.

An example of the shape of the 3D quasiperiodic solution is displayed, in
the projection onto plane $\left( x,y\right) $, in Fig. \ref{fig7}(b).
Additional examples can be found in Ref. \cite{Komarova}.

Besides these solutions, Eqs. (\ref{FWM}) give rise to another quasiperiodic
state, with $\varphi _{1}+\varphi _{2}-\varphi _{3}-\varphi _{4}=0$ and $%
\left\vert u_{l}\right\vert ^{2}=\gamma _{0}/9$ (cf. Eqs. (\ref{pi}) and (%
\ref{1/5})), but it is unstable. Also exist but are unstable two-mode
solutions, e.g., ones with $\left\vert u_{1,2}\right\vert ^{2}=\gamma
_{0}/3,u_{3,4}=0$ \cite{Komarova}.

\section{Conclusion}

The aim of this article is to present a concise overview of two important
topics in the theory of pattern formation in nonlinear dissipative media,
\textit{viz}., DWs (domain walls) and QP (quasiperiodic) patterns. The
topics are selected as those important contributions to which were made in
works of Prof. Mikhail Tribelsky. Most results collected in this article may
be considered as rather \textquotedblleft old" ones, as they had been
published ca. 30-35 \cite%
{Trib-DW,Trib-quasi,early,CQ-first,Burgers,Fauve,Alik,Steinberg,optical-DW,CGL-DW,Cross,Pomeau,MNT-quasi}
or 20 \cite{Kolodner,Komarova,Rotstein,Martin,Poland} years ago.
Nevertheless, these results remain relevant in the context of ongoing
theoretical and experimental studies in the ever expanding pattern-formation
research area. This conclusion is upheld by the fact that the present
article includes a few novel exact analytical results, obtained as a
relevant addition to the old theoretical findings concerning the DWs in
systems of coupled real GL (Ginzburg-Landau) equations \cite{new}. The new
results, represented by Eqs. (\ref{exact2}), (\ref{r2=0})-(\ref{xi}), (\ref%
{exact-v})-(\ref{DD}), and (\ref{exact3})-(\ref{sgn}) produce exact
solutions for symmetric DWs in the system of real GL equations including
linear mixing between the components, the solution for strongly asymmetric
DWs in the case when the diffusion term is present only in one GL equation,
the three-component composite state including the DW in two components and a
bright soliton in the third one, and the particular exact solution for DWs
between waves governed by the real GL equations including group-velocity
terms with opposite signs.

The significance of the results presented in this brief review is enhanced
by the fact that essentially the same coupled equations describe patterns of
the DW and QP types not only in thermal convection, but also in nonlinear
optics, BEC, and other physical systems. In particular, the pattern
formation in BEC of cesium atoms under the action of a temporally-periodic
modulation of the nonlinearity (imposed by means of the Feshbach resonance
\cite{Feshbach}), similar to the Faraday instability, was recently
experimentally demonstrated and theoretically modeled in the framework of
amplitude equations similar to Eqs. (\ref{gamma0}) in Ref. \cite{Chin}.
Another novel realization of the pattern formation was proposed for a driven
dissipative Bose-Hubbard lattice, that can be implemented in superconducting
circuit arrays \cite{Shanghai}.

It is expected that theoretical and experimental studies along the
directions outlined in this article have a potential for further
development, which will make it possible to add new findings to the
above-mentioned well-established results.

\section*{Acknowledgments}

First of all, I would like to thank Mikhail Tribelsky for the collaboration,
established long ago, which had produced essential results summarized in
this article. I also thank other colleagues in collaboration with whom these
results were obtained: Alexander Nepomnyashchy, Jerry Moloney, Alan Newell,
Natalia Komarova, Martin Van Hecke, and Horacio Rotstein. I appreciate 
a discussion of the topic of domain walls with Dmitry Pelinovsky, and the
help of Zhaopin Chen in producing Figs. \ref{fig_extra1} and \ref{fig_extra2}.

As one of editors of the Special Issue of journal Physics (published by
MDPI), dedicated to the celebration of the 70th birthday of Mikhail
Tribelsky, for which this article was written, I thank two other editors of
the Special Issue, Andrey Miroshnichenko and Fernando Moreno, for very
efficient collaboration.

This work was supported, in part, by Israel Science Foundation through grant
No. 1286/17.


\begin{thebibliography}{999}
\bibitem{name} Name known as
Mikhail/Michael/Michel/Miguel/Michele/Michal/Mikael/Mikkel/Mitxel/... is
considered as the oldest masculine name\ used in modern languages. The
meaning of its original form in Hebrew is ``who (is) like El (God)?", which
implies a response ``no one can be compared to God".
\end{thebibliography}
\end{document}